\documentclass[%
 reprint,
 amsmath,amssymb,
 aps,
]{revtex4-2}

\usepackage{graphicx}
\usepackage{dcolumn}
\usepackage{bm}

\usepackage[version=4]{mhchem}
\usepackage{siunitx}
\usepackage{soul}
\usepackage{xcolor}
\DeclareSIUnit\Molar{M}

\definecolor{mygray}{gray}{0.6}

\begin{document}

\preprint{APS/123-QED}

\title{Prevalence of mutualism in a simple model of microbial co-evolution}

\author{Luciano Stucchi}
\affiliation{Universidad del Pac\'{\i}fico, Lima, Peru}
\affiliation{Group of Complex Systems, Universidad Polit\'ecnica de Madrid, Madrid, Spain}

\author{Javier Galeano}
\author{Juan Manuel Pastor}
\affiliation{
 Group of Complex Systems, Universidad Polit\'ecnica de Madrid, Madrid, Spain
}%

\author{Jose Mar\'{\i}a Iriondo}
\affiliation{
 Biodiversity and Conservation Area, ESCET, Universidad Rey Juan Carlos, Madrid, Spain
}%

\author{Jos\'e A. Cuesta}
\affiliation{%
 Grupo Interdisciplinar de Sistemas Complejos (GISC), Madrid, Spain
}
\affiliation{%
 Dept.~Mathematics, Universidad Carlos III de Madrid, Legan\'es, Madrid, Spain
}
\affiliation{%
 Instituto de Biocomputaci\'on y F\'isica de Sistemas Complejos (BIFI),\\
 Universidad de Zaragoza, Zaragoza, Spain
}

\date{\today}

\begin{abstract}
Evolutionary transitions among ecological interactions are widely known, although their detailed dynamics remain absent for most population models. Adaptive dynamics has been used to illustrate how the parameters of population models might shift through evolution, but within an ecological regime. Here we use adaptive dynamics combined with a generalised logistic model of population dynamics to show that transitions of ecological interactions might appear as a consequence of evolution. To this purpose we introduce a two-microbial toy model in which population parameters are determined by a bookkeeping of resources taken from (and excreted to) the environment, as well as from the byproducts of the other species. Despite its simplicity, this model exhibits all kinds of potential ecological transitions, some of which resemble those found in nature. Overall, the model shows a clear trend toward the emergence of mutualism.
\end{abstract}

\maketitle


\section{\label{sec:Intro}Introduction}

In 1966, Jeon and Lorch were conducting experiments with a population of \textit{Amoeba proteus}. One of the strains got infected by X-bacteria, a Gram-negative, rod-shaped bacteria related to \emph{Legionella} sp. and \emph{Pseudomonas} sp. (for which the name \emph{Candidatus Legionella jeonii} sp. nov. was later proposed \citep{park:2004}). The few survivors to the infection retained the bacteria as parasites \citep{jeon:1967}. Over the course of a few generations though, they became endosymbionts, providing amoebas heatshock protection \citep{jeon:1992,jeon:1995}.

Arguably, this is one of the most spectacular cases of an evolutionary transition in an ecosystem from an antagonistic interaction to an obligate mutualism, mainly because it was witnessed in real time. However, other similar transitions are well documented in the literature from different sources. For instance, phylogenetic data collected from one of the best-known mutualistic systems---plants and pollinators---dating back 90 million years, allowed \cite{machado:2001} to show that the wasp \textit{Ceratosolen galili}, which belongs to a genus of active pollinating species, not only does not pollinate anymore but has become a parasitic species of figs.

Likewise, performing phylogenetic analysis on 15 species of aphids of the genus \textit{Chaitophorus}, \cite{shingleton:2002} concluded that the relationship between ants and aphids may have changed at least five times during the history of their life, thus going through all shades between mutualism and antagonism. There is no consensus whatsoever on whether the ancestral mutualistic relationship between these species was facultative or obligatory, but it seems reasonable to think that, before these species developed any mutualistic relationship, ants predated on aphids because this behaviour is still observed in all ant species in the appropriate environmental conditions \cite{sakata:1994,stadler:2005,offenberg:2001}.

The above-mentioned examples are only a few well-studied ecological systems that show transitions between different types of ecological interactions. In all of them, the evolutionary nature of these transitions has been established either through experiments or using phylogenetic analyses \citep{sachs:2006}, but their presence suggests that many more may have occurred in nature. This fact raises many questions: Are these evolutionary transitions a common phenomenon or a rarity? Why do ecological interactions move in one direction or another? Would it be possible to predict in which direction a type of interaction will move, if at all?

One way to explore the answers to some of these questions is to design mathematical models that include two different time scales: a short one that accounts for the usual population dynamics, and a long one that includes Darwinian evolution. The latter can be accomplished by applying adaptive dynamics to the parameters of the population model \cite{dieckmann:1996,dercole:2008,doebeli:2011}. However, the former is more problematic because standard population models deal with different types of interactions in very different ways.   

When it comes to modelling competition and predation, Lotka-Volterra population equations are a very convenient choice \cite{lotka:1925,volterra:1926,volterra:1928}. Despite their simplicity, these equations yield a rich set of biological predictions, to the point that they can be used as a prototype for more realistic models as well as a tool for interpreting complex observations. However, the Lotka-Volterra model meets serious difficulties to accommodate mutualism because positive interactions between species induce spurious feed-back loops that may drive populations out of control \cite{may:1981}. The addition of Holling-type functional responses \cite{wright:1989} is the usual way to control an unbounded growth of the populations, at the expense of rendering the model analytically intractable. Recently though, a new population model has been introduced in which both, species' intrinsic growth and interspecific interactions, are limited via logistic terms \cite{garcia-algarra:2014,stucchi:2020}. The resulting equations are amenable to analytic treatment and, at the same time, provide sensible results whether the interactions are of antagonistic or mutualistic nature. These two nice features make this model particularly suitable to study the evolution of ecological interactions.

Variation in ecological interactions may take place as a result of multiple factors \cite{thompson:1988}. For instance, the outcomes of ecological interactions may depend on the age and/or life cycle stage of the individuals of the interacting species. Similarly, phenotypic differences between the individuals of each of the interacting species, may alter the impact of one species on the other. These differences can result from genotypic diversity in the interacting populations and/or phenotypic plasticity. Environmental conditions might also determine how members of two populations interact with each other, allowing transitions from mutualism to competition or even extinction \cite{hoek:2016}. In order to focus on the general patterns derived from population dynamics and evolutionary adaptation, we have chosen a basic microbial system in which there are no differences among the individuals of each of the interacting species as a result of age or life cycle stage. Furthermore, there is one single genotype per interacting species and the environmental conditions are fixed. Therefore, all individuals of each interacting species have the same phenotype. Also for simplicity we will assume that mutations are so rare that they either get quickly fixed in the whole population or disappear before a new one occurs. In this regime, a suitable theoretical tool to study evolution is adaptive dynamics \cite{dieckmann:1996,dercole:2008,doebeli:2011}.

Accordingly, our toy model consists of two microbial species that interact via the resources they consume and excrete. They compete for resources from the environment---when both of them use the same resource---but they can also cross-feed from resources excreted by the other species. This interplay between resource consumption and excretion will provide specific functional forms for the parameters of \citeauthor{stucchi:2020}'s population equations, automatically yielding natural trade-offs between these parameters \cite{tilman:2004}. Adaptive dynamics will then introduce the Darwinian mechanism for the evolution of the parameters of the model.

Simplistic as it may be, this model provides sufficient complexity and flexibility as to show a great diversity of evolutionary transitions. It reveals a global trend toward the appearance of mutualistic interactions from initially competitive or antagonistic scenarios, and exhibits evolutionary pathways akin to some documented in the literature. 

In what follows we provide a detailed account of the model for two species, starting with a brief description of  \citeauthor{stucchi:2020}'s ecological model and further connecting its phenomenological parameters with the microscopic interactions between the species. The equations of the adaptive dynamic for this model are developed in the Appendix~\ref{app1}. We end by providing and analysing the results of extensive numerical simulations.

\section{A toy model for two interacting microbial species}

We will consider two microbial species whose populations evolve in time according to the \emph{generalised logistic model} \citep{stucchi:2020}, briefly described in Eq.~\eqref{eq:logistic}. This model is suitable for our purpose because it allows for all kinds of ecological interactions (whether beneficial or detrimental) and, at the same time, populations are always limited by carrying capacities. Generalized Lotka-Volterra equations lack this property, so that they end up having difficulties in handling mutualism---hence introducing a spurious bias toward competing interactions.

If we specialize this model for just two species, and ignore intraspecific cooperation or direct competition (i.e., $b_{ii}=0$), the two equations that describe this minimal ecological community are
\begin{equation}
\begin{split}
\dot{N_1} &=N_1(r_1-a_1N_1)+N_1(1-c_1N_1)b_{12}N_2, \\
\dot{N_2} &=N_2(r_2-a_2N_2)+N_2(1-c_2N_2)b_{21}N_1.
\end{split}
\label{eq:2population}
\end{equation}
where the constants $r_i$ are the intrinsic growth rates, the $a_i$ account for the effect of intraspecific competitions, the $b_{ij}$ are the interspecific interaction coefficients, and the $c_i$ are coefficients to saturate the effect of interspecific interactions. This system of differential equations is able to describe every kind of ecological interaction. Table~\ref{tab:rb} shows the kind of relationships of the species with the environment (through the signs of the intrinsic growth rates $r_i$) and between them (through the signs of the interaction constants $b_{ij}$) which correspond to the most usual ecological scenarios. Depending on the signs of $b_{ij}$ the interaction between both species can be \emph{mutualistic} (both positive), \emph{antagonistic} (one positive and one negative, also called \emph{predation} or \emph{parasitism,} depending on the context), or \emph{competitive} (both negative). On the other hand, the relation with the environment determines whether mutualism is \emph{facultative} ($r_i>0$) or \emph{obligate} ($r_i<0$). The case of \emph{predation/parasitism} ($b_{ij}<0$, $b_{ji}>0$, i.e.~species $j$ benefits at the expense of species $i$) requires the prey or parasitised species to obtain resources from the environment ($r_j>0$) for the system to be sustainable. When $b_{ij}\approx 0$, this sort of interaction is usually referred to as \emph{commensalism}. Finally, \emph{competition} ($b_{ij}<0$, $b_{ji}<0$) occurs when both species are a hindrance to each other (which of course needs $r_i>0$ and $r_j>0$ for the system to be sustainable).

\begin{table*}[t]
\caption{\label{tab:rb}Signs of $r_i$ and $b_{ij}$ that describe typical ecological interactions.}
\begin{ruledtabular}
\begin{tabular}{cccclc}
$r_i$ & $r_j$ & $b_{ij}$ & $b_{ji}$ & \multicolumn{1}{c}{\emph{type of interaction}} & abbreviation\\
\colrule
$+$ & $+$ & $+$ & $+$ & facultative-facultative mutualism & FFM \\
$+$ & $-$ & $+$ & $+$ & obligate-facultative mutualism & OFM \\
$-$ & $-$ & $+$ & $+$ & obligate-obligate mutualism & OOM \\
$+$ & $+$ & $-$ & $-$ & competition & CMP \\
$+$ & $+$ & $-$ & $+$ & facultative predator-prey / parasitism & FPP \\
$+$ & $-$ & $-$ & $+$ & obligate predator-prey / parasitism & OPP \\
$+$ & any & $\sim0$ & $+$ & commensalism & COM \\
\end{tabular}
\end{ruledtabular}
\end{table*}

\begin{figure}
\includegraphics[width=0.7\hsize]{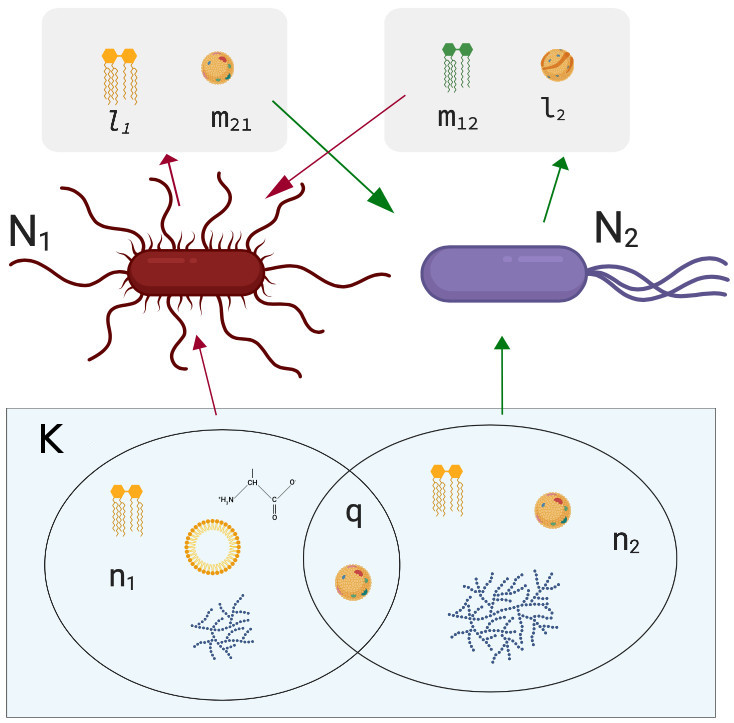}
\caption{Sketch of the model for two microbial species. Figures of microbes and molecules are purely symbolic, included only for the sake of illustration. They are meant to represent two species taking different resources from a pool and interchanging residues. Image created with \cite{Bio}.}
\label{fig:bacteria}
\end{figure}

Parameters $r_i$ and $b_{ij}$ are purely phenomenological. However, in a real system, they depend on the specific ways in which the species relate with the environment and with each other. Because of this, the parameters are not independent. However, to specify how they are connected, we need to delve into the details of the system we want to model. This is a necessary step to construct an evolutionary model of an ecosystem \cite{dieckmann:1996}.

Let us assume that the two microbial species described by Eq.~\eqref{eq:2population} struggle to survive in a fixed environment (described by the parameters $a_i$ and $c_i$), with a given number $K$ of available resources. Each species consumes $n_i$ of these $K$ resources, $q$ of which are common to both species (shared resources). As a consequence of internal metabolic reactions, species $i$ produces $m_i$ byproducts, $m_{ji}$ of which are useful to the other species $j$ and $\ell_i=m_i-m_{ji}$ are not (we do not care about metabolic waste in this model). Even though $m_i$ depends on the details of the metabolism of species $i$, for the sake of simplicity we can just assume that it is proportional to the number of consumed resources, i.e.
\begin{equation}
m_1=\gamma(n_1+m_{12}), \qquad m_2=\gamma(n_2+m_{21}),
\label{eq:m1m2}
\end{equation}
where $\gamma$ quantifies \emph{metabolic efficiency}. A fraction of species $i$'s metabolic byproducts will be used by species $j$; hence,
\begin{equation}
m_{21}=\alpha_2(n_1+m_{12}), \qquad m_{12}=\alpha_1(n_2+m_{21}),
\label{eq:m12m21}
\end{equation}
where $\alpha_j$ ($0<\alpha_j<\gamma$) measures \emph{cross-feeding efficiency}---parameter $\alpha_i$ determines the benefit that species $i$ obtains from species $j$---so it directly influences coefficient $b_{ij}$ in Eq.~\eqref{eq:2population}. In this toy model we assume that these cross-feeding efficiencies can evolve, which makes the coefficients in Eq.~\eqref{eq:2population} evolve accordingly.

Equation~\eqref{eq:m12m21} is a linear system whose solution is
\begin{equation}
m_{12}=\frac{\alpha_1(n_2+\alpha_2n_1)}{1-\alpha_1\alpha_2}, \qquad
m_{21}=\frac{\alpha_2(n_1+\alpha_1n_2)}{1-\alpha_1\alpha_2},
\label{eq:m12m21final}
\end{equation}
and substituting in Eq.~\eqref{eq:m1m2},
\begin{equation}
m_1=\frac{\gamma(n_1+\alpha_1n_2)}{1-\alpha_1\alpha_2}, \qquad
m_2=\frac{\gamma(n_2+\alpha_2n_1)}{1-\alpha_1\alpha_2}.
\label{eq:m1m2final}
\end{equation}
These equations express the total number of metabolic residues [c.f.~Eq.~\eqref{eq:m1m2final}] and the number of beneficial byproducts [c.f.~Eq.~\eqref{eq:m12m21final}] as a function of the number of resources.

 
Finally, we need to connect the demographic parameters with this flux of resources and byproducts. In principle, if resources are available in limited amounts, microbes would fare better the more resources they have, whereas metabolic byproducts are costly. It has been argued though that if resources are abundant becoming a specialist could be advantageous over being a generalist \cite{schwartz:1998}. The reason is that diversifying mechanisms to use different resources has a cost in fitness. For the sake of simplicity, we will stick to the first scenario. Thus, we posit that the growth rate increases proportional to the amount of consumed resources and decreases with the amount of metabolic byproducts, i.e.
\begin{equation}
\begin{split}
r_1 &=\frac{r_{10}}{K}(n_1-m_1)\\
&=\frac{r_{10}}{1-\alpha_1\alpha_2}\big[(1-\gamma-\alpha_1\alpha_2)u_1-\gamma\alpha_1u_2\big], \\
r_2 &=\frac{r_{20}}{K}(n_2-m_2)\\
&=\frac{r_{20}}{1-\alpha_1\alpha_2}\big[(1-\gamma-\alpha_1\alpha_2)u_2-\gamma\alpha_2u_1\big],
\end{split}
\label{eq:ripars}
\end{equation}
denoting $u_i\equiv n_i/K$. Notice that the factor $K$ is introduced for convenience, so as to express everything in terms of the re-scaled variables $0\leqslant u_i\leqslant 1$. (See \cite{suppmat} for a list of the parameters and symbols used in this model.)

Likewise, the interaction coefficients $b_{ij}$ increase with $m_{ij}$, the number of byproducts of species $j$ that are useful to species $i$, and decrease with the competition for the shared resources $q$. Hence,
\begin{equation}
\begin{split}
b_{12} &=\frac{b_{10}}{K}(m_{12}-q)\\
&=b_{10}\left[\frac{\alpha_1}{1-\alpha_1\alpha_2}(\alpha_2u_1+u_2)-w\right], \\
b_{21} &=\frac{b_{20}}{K}(m_{21}-q)\\
&=b_{20}\left[\frac{\alpha_2}{1-\alpha_1\alpha_2}(\alpha_1u_2+u_1)-w\right],
\end{split}
\label{eq:bijpars}
\end{equation}
where we have denoted $w\equiv q/K$ ($0\leqslant w\leqslant\min(u_1,u_2)$).

Parameters $r_{i0}$ and $b_{i0}$ are simple dimensional constants that also set the time scale of the differential equations.

Thus, after these simplifying assumptions, we end up with a model in which the demographic parameters are expressed in terms of the number of consumed resources, $n_1$, $n_2$, the resource sharing $q$, and the cross-feeding efficiencies $\alpha_1$, $\alpha_2$. A sketch of the model, where all this interactions are summarized, is shown in Figure~\ref{fig:bacteria}. The pool of resources is represented by the rectangle $K$ in blue, where $n_1$ and $n_2$ are the sets of resources useful for species $1$ and $2$, respectively, and $q$ is the set of shared resources. Residues excreted by each species are in the upper gray rectangles. Note that the shape and color of microbes are meaningless, and so are the molecules depicted. They are used just for illustration purposes.

\section{Adaptive dynamics}

According to Eq.~\eqref{eq:2population}, the dynamics of species $i$'s population fits the pattern
\begin{equation}
\dot{N}_i=N_if_i(\Omega,N_1,N_2),
\label{eq:fitness}
\end{equation}

where $\Omega$ denotes the set of parameters $\{u_1,u_2,w,\alpha_1,\alpha_2\}$. The function $f_i(\Omega,N_1,N_2)$ describes the per-capita growth rate, or \emph{fitness,} of species $i$, a magnitude that depends on the population growth parameters as well as the population sizes of the two species involved. Any steady state of the community, $N_i=N_i^*$, is defined by the equations
\begin{equation}
f_i(\Omega,N_1^*,N_2^*)=0.
\end{equation}
When a community is in a given steady state, a mutant may appear in one of the populations. If the mutant belongs to species $i$, its fitness will be a function $\bar{f}_i(\Omega',\Omega,N_1,N_2)$, where the prime parameters are those of the mutant. If this fitness is negative the mutant will go extinct, otherwise it can increase its frequency in the population and replace the original genotype.

Adaptive dynamics (AD) is a method to exploit this idea to devise a set of differential equations for the demographic parameters. Under the assumption that the parameters of the mutant are a small perturbation of those of the original genotype, \cite{dieckmann:1996} derived these differential equations from the master equation of the underlying stochastic mutation-selection process.

The derivation of these so-called canonical equations of AD implicitly assumes that the demographic parameters can vary independently of each other, and that mutations have the same probability to increase or decrease these parameters by the same small amount. Neither of these two conditions are met in our model under the scheme of mutations of this system (increasing or decreasing the number of resources). One kind of mutations amounts to changing the status of a randomly chosen external resource, i.e., adding the resource if it is new, or dropping it if the species was already using it. This means that $n_i\to n_i+1$ with a probability proportional to the number of new resources, and $n_i\to n_i-1$ with a probability proportional to the number of resources already in use. But at the same time, $q$, the number of common resources to both species, can increase, decrease, or remain unchanged. Table~\ref{tab:transitions} summarises all mutational scenarios and their corresponding probabilities.

\begin{table*}[t]
\caption{\label{tab:transitions}Set of mutations in the system that change the number of external resources and their corresponding probabilities. Here $n_i$ are the number of external resources consumed by species $i$, $q$ the amount of them common to both species, and $u_i\equiv n_i/K$, $w\equiv q/K$.}
\begin{ruledtabular}
\begin{tabular}{llllllll}
\multicolumn{4}{c}{initial state: $n_1$, $q$} &
\multicolumn{4}{c}{initial state: $n_2$, $q$} \\
\colrule
\multicolumn{2}{l}{mutated state} & & probability & \multicolumn{2}{l}{mutated state} & & probability \\
\colrule
$n_1+1$, & $q+1$ & & $u_2-w$ &
$n_2+1$, & $q+1$ & & $u_1-w$ \\
$n_1+1$, & $q$ & & $1-u_1-u_2+w$ &
$n_2+1$, & $q$ & & $1-u_1-u_2+w$ \\
$n_1-1$, & $q$ & & $u_1-w$ &
$n_2-1$, & $q$ & & $u_2-w$ \\
$n_1-1$, & $q-1$ & & $w$ &
$n_2-1$, & $q-1$ & & $w$ \\
\end{tabular}
\end{ruledtabular}
\end{table*}

The second kind of mutations will change $m_{ij}$, the amount of byproducts that species $i$ take from species $j$, by $\pm 1$. As $\alpha_i$ is a multiplicative factor,
\begin{equation}
\frac{\delta m_{ij}}{m_{ij}}=\frac{\delta\alpha_i}{\alpha_i},
\end{equation}
so $\delta m_{ij}=\pm 1$ implies
\begin{equation}
\delta\alpha_i=\frac{\pm\alpha_i}{m_{ij}},
\end{equation}
or, using Eq.~\eqref{eq:m12m21final},
\begin{equation}
\begin{split}
\delta\bar\alpha_1\equiv\frac{u_2+\alpha_2u_1}{1-\alpha_1\alpha_2}\delta\alpha_1=\pm\frac{1}{K}, \\ 
\delta\bar\alpha_2\equiv\frac{u_1+\alpha_1u_2}{1-\alpha_1\alpha_2}\delta\alpha_2=\pm\frac{1}{K}.
\end{split}
\label{eq:mutationalpha}
\end{equation}
On the other hand, $\delta\alpha_i<0$ with probability $m_{ij}/m_i=\alpha_i/\gamma$ and $\delta\alpha_i>0$ with probability $\ell_i/m_i=(\gamma-\alpha_i)/\gamma$. (See \cite{suppmat} for a list of variables and symbols used in this model .)

In summary, both species can experience two kinds of mutations. The first one changes the number of resources the species consumes, that is, it may add one new resource or dispose of an old one, hence increasing or decreasing $n_i$ one unit. Notice that this added or removed resource may or not be shared with the other species. The second kind of mutation is related with its ability to use byproducts excreted to the environment by the other species---an ability accounted for by the parameter $\alpha_i$.

Reconstructing the procedure of \cite{dieckmann:1996} to obtain the canonical equation of AD, we arrive at a system of differential equations for $u_1$, $u_2$, $w$, $\alpha_1$, $\alpha_2$ (c.f.~Eq.~\eqref{eq:evolut_u1_u2_w} and Eq.~\eqref{eq:evolut_alpha} in Appendix~\ref{app1}; see the Supplementary Material for a more detailed derivation of the equations). Notice that the stationary populations $N_1^*$, $N_2^*$ appear explicitly in the equations, and that these populations depend in turn on the parameters $u_1$, $u_2$, $w$, $\alpha_1$, $\alpha_2$.

\section{Numerical simulations}

As the differential system \eqref{eq:evolut_u1_u2_w} is very difficult to discuss analytically, we have performed numerical simulations of the time evolution of the system for different initial conditions and sets of parameters. Two timescales are involved here: the population changes in a fast timescale, whereas the parameters evolve slowly. The evolutionary parameters are obtained by integrating Eqs.~\eqref{eq:evolut_u1_u2_w} and \eqref{eq:evolut_alpha}. We do that using a fourth-order Runge-Kutta method (RK4) to obtain the parameters $\{u_1(t),u_2(t),w(t),\alpha_1(t),\alpha_2(t)\}$ in evolutionary time. Note that RK4 needs the slope of the function at 4 different times and that slope depends explicitly on the stationary populations $N_1^*$, $N_2^*$ through Eq.\eqref{eq:N1N2}, so at each of those times we need to calculate the stationary populations---which depend, in turn, on the evolutionary parameters $\{u_1,u_2,w,\alpha_1,\alpha_2\}$ at those same times.

All simulations have been run with a time-step $\Delta t=10^{-5}$ and were stopped as soon as they reached a stationary state---when the evolutionary parameters did not change more than an error tolerance $10^{-4}$ over $10^4$ time steps---or after $10^7$ times steps.

\section{Evolutionary attractors}

Using different sets of parameters and initial conditions, all the evolutionary attractors that we have observed fit within just a few patterns. The triad of parameters $(u_1,u_2,w)$ (fraction of resources consumed by each species and fraction of shared resources) is always found to end up in one of the three forms $(1,u,u)$,  $(1-u,u,0)$, or $\big(1-u,u,\min(1-u,u)\big)$, with $0\le u\le 1$.
If $(1,u,u)$ is reached, the evolutionary attractor turns out to be either competition or antagonism. In particular, in the case $(1,1,1)$ the attractor is always competition. As expected, when the final state is mutualism or commensalism no resources are shared ($w=0$) (there is no need for competing for resources), so when $(1-u,u,0)$ is reached the system ends up being mutualistic.

With respect to the cross-feeding efficiencies, $\alpha_i$, the system generally evolves towards extreme values of these parameters ($\alpha_i=0$ or $\alpha_i=\gamma$). Only when the system evolves to competition can some intermediate values be found. As expected, cross-feeding efficiencies reach their maximum ($\alpha_1=\alpha_2=\gamma$) for mutualism and their minimum ($\alpha_1=\alpha_2=0$) mainly for competition.

It is worth mentioning that all kinds of mutualisms are found to be evolutionary attractors for some initial conditions. Another relevant observation is that evolution sometimes drives ecosystems to extinction. This is no longer a surprise because it is a result that has already been empirically observed (e.g.~in viruses \cite{turner:1999}), but the idea that evolution can degrade ecosystems would have shocked evolutionists of the 19th and early 20th century because it contradicts the notion of evolution as a sort of `optimiser'.

\section{Evolutionary transitions between types of ecological interactions}

In order to illustrate the kind of evolutionary transitions between types of ecological interactions that this system can produce, we performed an exhaustive exploration of parameters. We fixed the environmental parameters $a_1=a_2=c_1=c_2=0.001$ and, without loss of generality, chose $r_{10}=1$ (this simply sets the evolutionary time scale). For the other species, we explored uniformly the interval $0<r_{20}\le 1$ and then zoomed in the region $0<r_{20}\ll r_{10}$ by sampling the interval $0<r_{20}\le 0.1$. For $b_{i0}$ we sampled uniformly the range $0\le b_{i0}\le 0.01$ and then zoomed in the interval $0\le b_{i0}\le 0.001$. Likewise, nine different, uniformly spaced values of the metabolic efficiency $\gamma$ within the range $0<\gamma<1$ were explored.

For each set of parameters we generated random initial conditions for $u_i$, $w$, and $\alpha_i$ within the ranges $0\le u_i \le 0.99$, $0\le w\le\min(u_1,u_2)$, and $0\le\alpha_i\le\gamma$, and kept only those that generated viable populations. Then we let each of these remaining cases evolve according to Eq.~\eqref{eq:evolut_u1_u2_w} and Eq.~\eqref{eq:evolut_alpha}. We recorded 1000 runs that resulted in viable populations for each set of parameters, discarding all initial conditions that led to no viable stationary populations but keeping track of those that eventually led to extinction. Notice that the number of resources or the mutation probability are only relevant to set the evolutionary time scale (see Appendix~\ref{app1}).

In order to catalogue the resulting evolutionary attractors we have followed the classification of Table~\ref{tab:rb}---considering a state as commensalist if one of the parameters $b_{ij}$ is positive and the other one is smaller than $10^{-8}$.

The results of this numerical study of the model are summarized in Fig.~\ref{fig:r1_bp01}, \ref{fig:rp1_bp01}, \ref{fig:r1_bp001}, and \ref{fig:rp1_bp001} (for the distribution of the initial and final states see the Supplementary Material). In what follows we describe in more detail the transitions from an initial ecological state to another one that we observed, depending on the choice of parameters $r_{i0}$ and $b_{i0}$.

\subsubsection{Parameters: $r_{20}\le 1,\ b_{i0}\le 0.01$}

\begin{figure}
    \centering
    \includegraphics[width=\hsize]{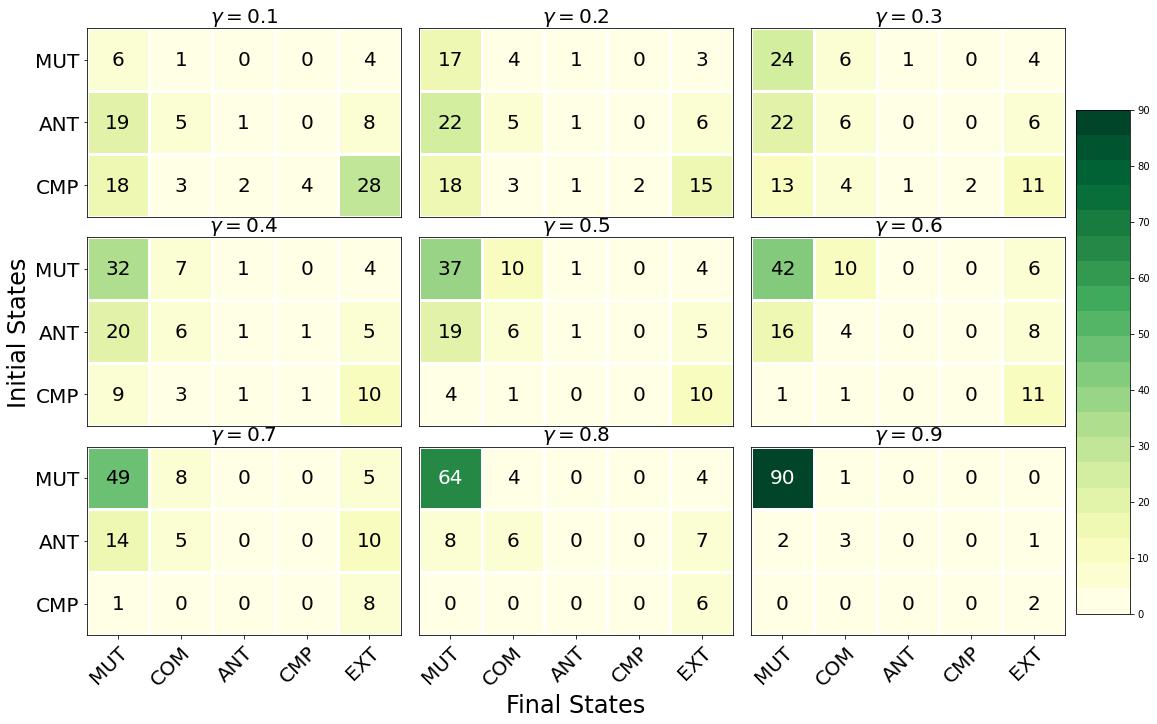}
    \caption{Evolutionary transitions for values of $\gamma$ ranging from $0.1$ to $0.9$. In rows, we show the initial ecological interaction states (mutualism (MUT), antagonism (ANT), and competition (CMP)), and in columns, the final ecological interaction states---with the addition of commensalism (COM) and extinction (EXT). Figures denote the probability of the corresponding transitions. Parameter values: $r_{20}\le 1$ and $b_{i0}\le 0.01$. For $\gamma\le 0.2$, the initial states are mainly ANT and CMP, and they mostly end up in MUT or EXT. For $\gamma\ge 0.3$, the initial states are mainly MUT and remain MUT.}
    \label{fig:r1_bp01}
\end{figure}


(See Fig.~\ref{fig:r1_bp01}.) For this choice of parameters most evolutionary pathways ended up in mutualism, from $\sim40\%$ to more than $90\%$ of the cases.
For lower values of $\gamma$, from $0.1\le\gamma\le0.3$, $\sim40\%$ of cases begin as antagonism or competition and extinction occurs in more than $20\%$ of all transitions. For higher values of $\gamma$, from $0.4\le\gamma\le0.7$, mutualism is the main initial state representing more than $30\%$ of the cases and extinction reduces to less than $20\%$. For $0.8 >\gamma$, mutualism accounts for $\sim90\%$ of the initial and final states, and extinction represent less than $16\%$ of the latter. In most cases, except for $\gamma=0.9$, commensalism accounts for $\sim10\%$ of the final states.






\subsubsection{Parameters: $r_{20}\le 0.1,\ b_{i0}\le 0.01$}

\begin{figure}
    \centering
    \includegraphics[width=\hsize]{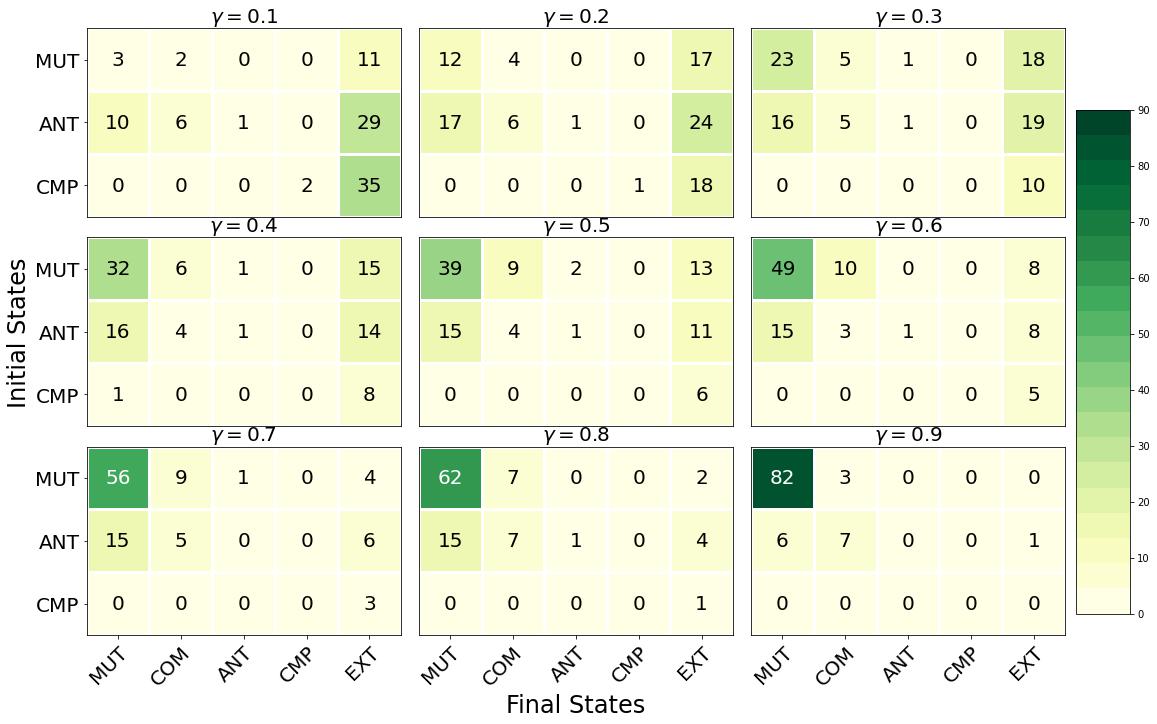}
    \caption{As Fig.~\ref{fig:r1_bp01}, for parameters values $r_{20}\le 0.1$ and $b_{i0}\le 0.01$. For $\gamma\le 0.2$, the initial states are mainly CMP and ANT, although they mostly end up in EXT. For $\gamma\ge 0.3$, the initial states are mainly MUT and remain MUT.}
    \label{fig:rp1_bp01}
\end{figure}


(See Fig.~\ref{fig:rp1_bp01}.) For this choice of parameters most evolutionary pathways ended up in mutualism or went extinct. Extinction accounts for more than $45\%$ of final states for $\gamma\le0.3$ reducing its importance for higher values until $\gamma=0.9$, where represents less than $2\%$. Mutualism begins being less than $15\%$ of all final states but ended up being almost $90\%$ of the cases. Commensalism remains representing $\sim10\%$ of the final states for all $\gamma$. Initial states are mainly antagonism and competition for lower values of $\gamma$, representing more than $80\%$ for $\gamma=0.1$ but turned to mutualism for their higher values, where it represents more than $80\%$ for $\gamma=0.9$.





\subsubsection{Parameters: $r_{20}\le 1,\ b_{i0}\le 0.001$}

\begin{figure}
    \centering
    \includegraphics[width=\hsize]{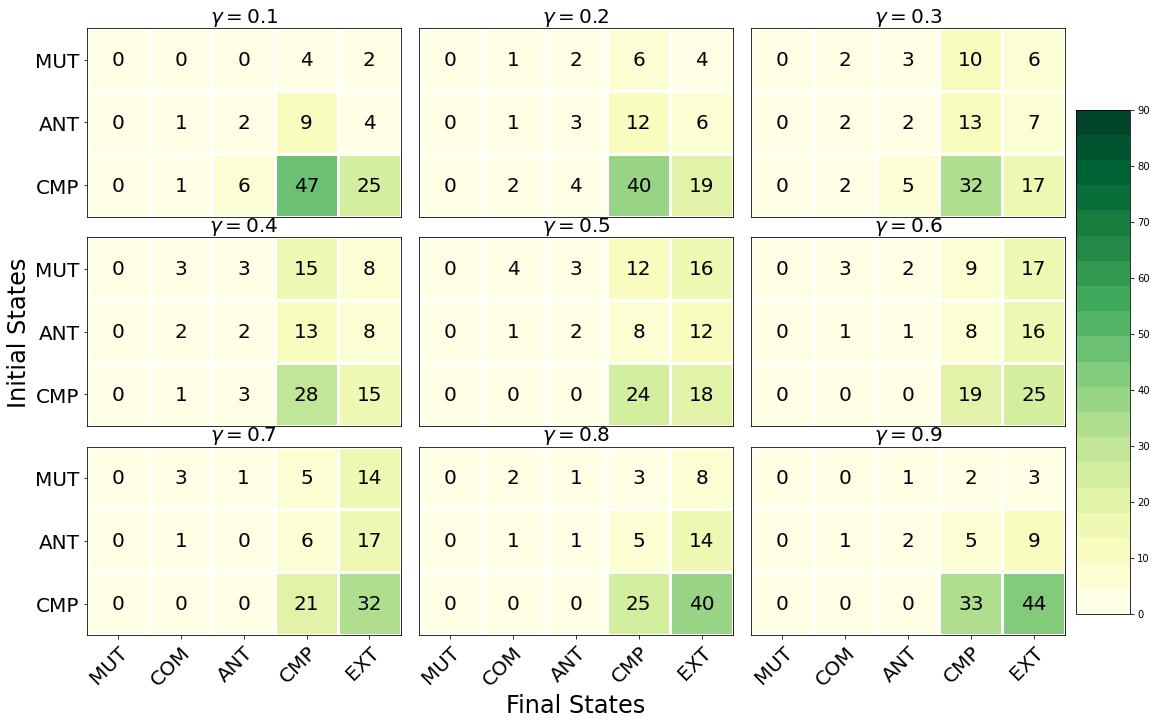}
    \caption{As Fig.~\ref{fig:r1_bp01}, for parameters values $r_{20}\le 1.0$ and $b_{i0}\le 0.001$. For $\gamma\le 0.3$, the initial states are mainly CMP, and they mostly stay as CMP. For $0.3\le\gamma\le 0.7$, communities can also begin as MUT or ANT. For $\gamma\ge 0.6$, most populations end up in CMP or EXT.}
    \label{fig:r1_bp001}
\end{figure}

(See Fig.~\ref{fig:r1_bp001}.) For this parameters the system evolved mainly to competition or went extinct. With such a small interaction parameters $b_{i0}$ mutualism is dramatically hindered---even in the cases where one third of the initial states were mutualistic (for $\gamma=0.5$). More than $40\%$ of the system started in competition and remained as such or went extinct---the proportion of which changed upon increasing $\gamma$ from $46.8\%$ competition vs. $24.8\%$ extinction, to $33.3\%$ competition vs. $44.2 \%$ extinction.

\subsubsection{Parameters: $r_{20}\le 0.1,\ b_{i0}\le 0.001$}

\begin{figure}
    \centering
    \includegraphics[width=\hsize]{rp1_b0p001_YlGn.jpg}
    \caption{As Fig.~\ref{fig:r1_bp01}, for parameters values $r_{20}\le 0.1$ and $b_{i0}\le 0.001$. For $\gamma\le 0.3$, the initial states are predominantly CMP, and they mostly end in EXT. For $\gamma\ge 0.4$, the initial states are predominantly MUT, although EXT continues to be the most common evolutionary fate. However, for $\gamma\ge 0.8$, COM and ANT become relatively common ending states.}
    \label{fig:rp1_bp001}
\end{figure}


(See Fig.~\ref{fig:rp1_bp001}.) As in the previous case, the system cannot evolve into a mutualistic state, but commensalism accounts for almost $10\%$ of the final states when $\gamma\le0.3$ and more than $15\%$ for $0.4\le\gamma$. Antagonism as well accounts for more than $30\%$ of the final cases when $\gamma=0.9$, being lower for $\gamma\le0.8$. Extinction is the main transition for all cases. Even though competition is the main initial state for lower values of $\gamma$, being more than $60\%$ of the initial cases for $\gamma=0.1$ and around $20\%$ for $0.5\le\gamma$, it shares the distribution with mutualism and antagonism when $\gamma$ increases, both of them being more than $20\%$ for all $0.2\le\gamma$.




\section{Transient states}

\begin{figure}
    \centering
    \includegraphics[width=\hsize]{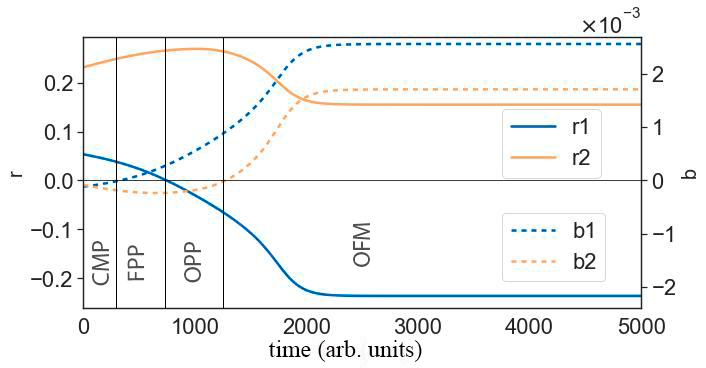}
    \caption{Evolutionary transition of the population parameters $r_i$ and $b_i$ of two species. The community starts in competition (CMP) and ends in obligate-facultative mutualism (OFM), going through intermediate states of facultative parasitism (FPP) and obligate parasitism (OPP). Initial parameters: $r_{01}=1$, $r_{02}=0.65727$, $b_{01}=0.00414$, $b_{02}=0.00447$, $u_1=0.17967$, $u_2=0.78500$, $w=0.10050$, $\alpha_1=0.08365$, $\alpha_2=0.32459$, $\gamma=0.5$.}
    \label{fig:trans1}
\end{figure}

\begin{figure}
    \centering
    \includegraphics[width=\hsize]{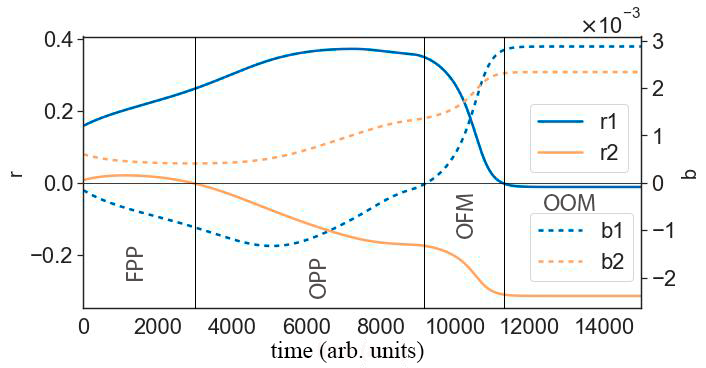}
    \caption{Evolutionary transition of the population parameters $r_i$ and $b_i$ of two species from facultative parasitism (FPP) to obligate-obligate mutualism (OOM), with intermediate states of obligate parasitism (OPP) and obligate-facultative mutualism (OFM). Initial parameters: $r_{01}=1.0$, $r_{02}=0.64396$, $b_{01}=0.00474$, $b_{02}= 0.00262$, $u_1=0.60097$, $u_2=0.62973$, $w=0.16897$, $\alpha_1=0.13282$, $\alpha_2=0.54215$ and $\gamma=0.6$.}
    \label{fig:trans2}
\end{figure}

\begin{figure}
    \centering
    \includegraphics[width=\hsize]{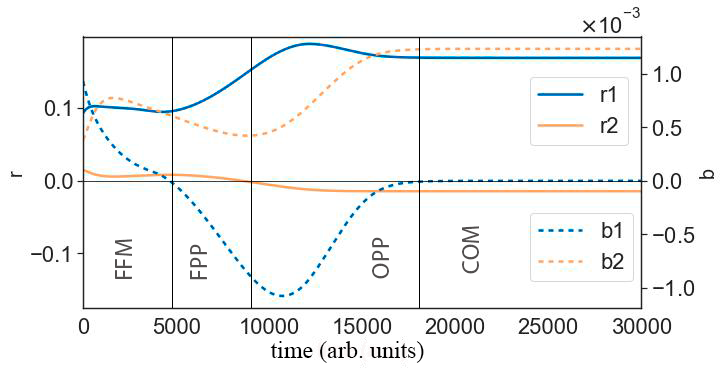}
    \caption{Evolutionary transition of the population parameters $r_i$ and $b_i$ of two species from facultative-facultative mutualism (FFM) to commensalism (COM), with intermediate states of facultative parasitism (FPP) and obligate parasitism (OPP). Initial parameters: $r_{01}=1.0$, $r_{02}= 0.09827$, $b_{01}=0.00463$, $b_{02}=0.00313$, $u_1=0.79063$, $u_2= 0.80318$, $w=0.00460$, $\alpha_1=0.22106$, $\alpha_2=0.12717$, and  $\gamma=0.7$.}
    \label{fig:trans4}
\end{figure}

\begin{figure}
    \centering
    \includegraphics[width=\hsize]{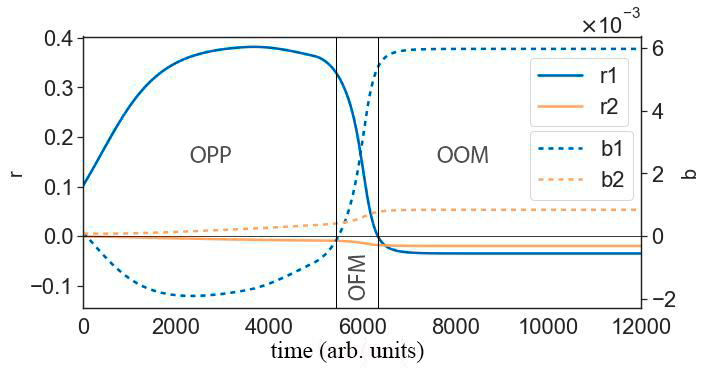}
    \caption{Evolutionary transition of the population parameters $r_i$ and $b_i$ of two species from obligate parasitism (OPP) to obligate-obligate mutualism (OOM), with an intermediate state of obligate-facultative mutualism (OFM). The initial OPP has a long stasis, wherear the intermediate OFM is much shorter in evolutionary scale. Initial parameters: $r_{01}=1$, $r_{02}=0.04230$, $b_{01}=0.00962$, $b_{02}=0.00096$, $u_1=0.45854$, $u_2=0.26628$, $w=0.12217$, $\alpha_1=0.28321$, $\alpha_2=0.35339$, and $\gamma=0.6$.}
    \label{fig:trans5}
\end{figure}

\begin{figure}
    \centering
    \includegraphics[width=\hsize]{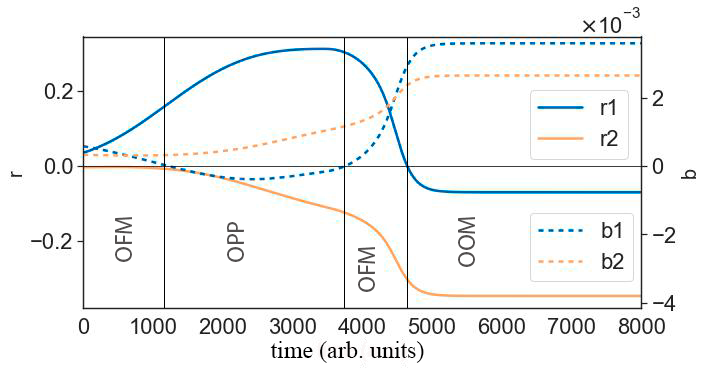}
    \caption{Evolutionary transition of the population parameters $r_i$ and $b_i$ of two species from obligate-facultative mutualism (OFM) to obligate-obligate mutualism (OOM), with an intermediate state of obligate parasitism (OPP) and obligate-facultative mutualism (OFM). Initial parameters: $r_{01}=1.0$, $r_{02}=0.80306$, $b_{01}=0.00561$, $b_{02}=0.00310$, $u_1=0.26642$, $u_2=0.17242$, $w=0.01485$, $\alpha_1=0.40480$, $\alpha_2=0.31492$, and $\gamma=0.6$.}
    \label{fig:trans6}
\end{figure}

\begin{figure}
    \centering
    \includegraphics[width=\hsize]{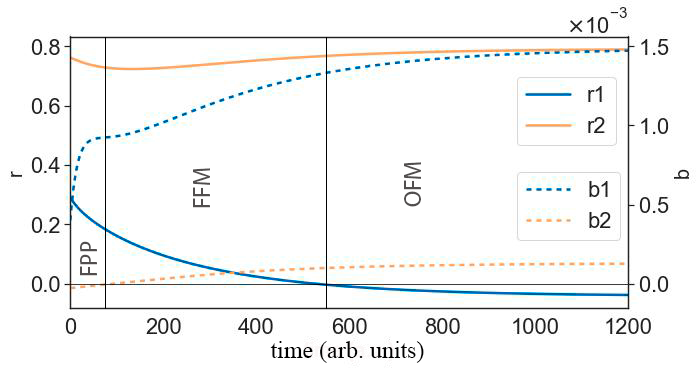}
    \caption{Evolutionary transition of the population parameters $r_i$ and $b_i$ of two species from facultative parasitism (FPP) to obligate-facultative mutualism (OFM), with an intermediate state of facultative-facultative mutualism (FFM). Initial parameters: $r_{01}=1.0$, $r_{02}=1.0$, $b_{01}=0.00713$, $b_{02}=0.00309$, $u_1=0.38958$, $u_2=0.96352$, $w=0.05325$, $\alpha_1=0.10848$, $\alpha_2=0.08866$, and $\gamma=0.2$.}
    \label{fig:trans7}
\end{figure}

Because of the rich evolutionary dynamics of this model, evolutionary transitions are not the only relevant feature to focus on. Particular sequences of transient ecological states along evolutionary pathways are as interesting---if not more so. In Figs.~\ref{fig:trans1}, \ref{fig:trans2}, \ref{fig:trans4}, \ref{fig:trans5}, \ref{fig:trans6}, and \ref{fig:trans7} we show the time evolution of the demographic parameters of just a few examples, chosen because of their peculiar sequence of intermediate transitions or because they resemble actual transitions observed in nature. (For the time evolution of the stationary populations $N_i$ and their evolutionary parameters, see the Supplementary Material).

Fig.~\ref{fig:trans1} shows two species with positive intrinsic growth rate that are initially in competition. Over time one of them starts parasitizing the other until it becomes dependent on it. Eventually, the other species `learns' to take advantage of the parasite and the relation ends up as a mutualism. Transitions from antagonistic relations to mutualism are well known to occur in microbial communities due to environmental pressure and phenotypic plasticity \cite{hoek:2016}. However, \cite{harcombe:2018} showed that mutualism arises as an evolutionary change in a controlled experiment involving \textit{Escherichia coli} and \textit{Salmonella enterica}. \textit{E. coli} strains went through genetic changes due to a mutation that led them from generating acetate---a costless byproduct that was useful to \textit{Salmonella}---to secreting the more useful galactose, even when this was a costly byproduct that reduced its intrinsic growth rate. This transition arose when \textit{Salmonella} strains were forced and selected to produce methionine, which was beneficial to \textit{E. coli}. Even though \textit{E. coli} mutant populations did not replaced the ancestral strains---as they were also benefited on this stage---this case shows how species must pay a cost in order to become mutualists.

In Fig.~\ref{fig:trans2} species 2 starts as a facultative parasite of species 1. Soon the parasite becomes dependent, and this situation remains like this for a while, until suddenly (in evolutionary terms) the relation evolves into a mutualism, facultative at first for species 1, but eventually obligate for both species.

A similar co-evolutionary pathway can be found in nature between Macrotermitinae (species 1) and fungi (species 2). According to \cite{nobre:2010} and \cite{aanen:2002}, both the fungi and the fungus-growing termites are obligate mutualists since they need each other in order to survive and reproduce. As stated by \cite{margulis:2002b}, it is plausible that the origin of the termite-fungi mutualistic relation was an infection (a specialized infestation) of the termites guts with fungi spores, which led them to defend themselves by domesticating the fungi, controlling and limiting their growth.

Fig.~\ref{fig:trans4} is an example of an opposite transition, in which one of the species of an initially mutualistic system evolves into a parasite of the other (first facultative, eventually obligate) which, over time, develops a commensalistic relation with the parasitized species.

Commensalism arises frequently in nature. For instance some algae and Ascomycota fungi do not form lichens, but descend of lichen-forming fungi ancestors. Non-lichenized fungi might obtain their nutrients acting as commensalists of lichen-forming fungi and algae \cite{lutzoni:2001}. Although the evolution of the lichen symbiosis is believed to have appeared multiple independent times \cite{divakar:2015}, no specific route for this formation has been described with certainty, to our knowledge.

There are many other cases of known mutualist relations that had become commensalistic. Zooxanthellae and octocorals form an ancestral facultative mutualistic relation where the dinoflagellates contribute to the energy budget of the invertebrates, by being host inside them. However, \cite{vanoppen:2005} showed that there is compelling evidence of a FFM$\rightarrow$COM transition, within the octocoral family Melithaeidae. Zooxanthellae and antheopleura form a similar relation and \cite{geller:2001} stated that many species of the sea anemones seem to have lost the mutualistic relationship with the algae. \textit{Gymnodinium} algae provide an energetic supply to many marine invertebrates, which in return act as protective hosts. \cite{wilcox:1998} show, through a molecular phylogenetic analysis, that \textit{Gymnodinium} is a genus that incorporate both mutualistic and independent species even though they all descend from a common symbiotic ancestor.

Similar cases can be found within microbial organisms. In genera \textit{Entamoeba} and \textit{Trypanosoma}, transitions from mutualism to what in our framework can be described as mutual commensalism have been reported. Clark and Roger \cite{clark:1995} reported that \textit{Entamoeba histolytica} shows evidence of mitochondrial relics, which might mean that those organelles, first acquired through endosymbiotic mechanisms, were eventually lost as a result of an evolutionary process. Hannaert et al.~\cite{hannaert:2003} also reported that two genera of Trypanosomatidae (\textit{Leishmania} and \textit{Trypanosoma}) acquired their plastids probably from mutualistic algae, and they also lost them leaving only a few residual genes. Such interpretations are allowing a better understanding of eukaryotic lineages since organisms formerly classified as divergent ancestors, like some archezoan protists \cite{clark:1995} or nematodes having and lacking their \textit{Wolbachia pipientis} simbionts \cite{casiraghi:2004}, are being understood now as much closer relatives. These studies show evolutionary transitions that are fundamentally different those shown in microbial studies due to environmental pressures \cite{hoek:2016} even where genetic manipulation is involved  \cite{lasarre:2017}.

In Fig.~\ref{fig:trans5} an initially parasitic relation, with a long period of stasis, eventually evolves into an interdependent mutualistic relation after crossing a brief period of facultative mutualism. Notice that the co-evolution of \emph{Legionella jeonii} and \emph{Amoeba proteus} described in the Introduction illustrates this kind of evolutionary pathway \cite{jeon:1967}.

Another example, Fig.~\ref{fig:trans6}, exhibits a case that begins and ends in a mutualistic state, but not before going through a period of parasitism. This case reveals that, even though some of the cases reported in the statistics of the previous section appear not to have undergone any transition whatsoever, they nevertheless can come across different intermediate states before reaching their evolutionary stable state.

As a last example, Fig.~\ref{fig:trans7} exhibits a case that begins in facultative parasitism and ends in a obligate-facultative mutualism, going through a period of facultative-facultative mutualism. This case is similar to the case of ants and aphids from several genera, whose relations have gone through all mutualistic and commensalistic nuances, as we pointed out in the Introduction.

Attine ants and fungi might have also undergone a similar pathway, since they form mutualistic relations that range from almost mutually obligatory to facultative---at least for the fungi. In particular, \cite{nobre:2010} indicate that the fungal symbionts of the higher attines depend almost exclusively on the ants---although occasionally they might reproduce sexually. However, \cite{schultz:2008} and \cite{currie:2003} also show that fungi are capable of living without the symbiotic relation with the ants.

Several hypotheses have been drawn to explain the evolutionary origin of the attine ant-fungus relation. According to \cite{mueller:2001}, even though it is widely accepted that fungi were part of the ancestral ant diet (which would constitute a predatory relation), neutral coexistence, where ants acted as accidental vehicles of fungal dispersion, may be a more viable explanation (in which case, the ancestral relation would be commensalistic, instead of predatory).

\section{Discussion}

Admittedly, the model we have proposed and analyzed in this paper is a far cry from any realistic description of microbial interactions. For instance, we have deliberately ignored any detail on the metabolic processes involved---which are determinant in deciding which products can or cannot be re-used---or kept the environmental conditions constant---hence neglecting any effect that a change in the environment might bring into the interactions \cite{thompson:1988,hernandez:1998}. For these and many other drastic simplifications we have assumed in its design, it is only fair to call our model a `toy model'. And yet this is precisely one of its virtues, because what this model makes clear is that very few assumptions about the way two species can interact lead to drastic changes in the ecological scenario. More and more, empirical evidence is showing that evolutionary transitions in the ecological interactions between species of a community are the norm, rather than the exception. The model proposed here provides a proof of concept in this sense, because it links natural microscopic interactions between microbes to subsequent changes in their ecological relationships.

It has been hypothesized that environmental fluctuations are one of the forces that might drive the evolution of microbial consumption and production capabilities. Our simple model does not dispute this point of view, but the fact that the random evolution of the interspecific interactions in a constant environment, subject to some trade-offs, is able to produce such a plethora of complex transitions between different ecological regimes, can only mean that the environment may not be their only driver. As a matter of fact, endosymbiosis is a process that occurs within a constant environment---and one that our simple model is able to capture. Certainly, environmental changes may trigger this evolutionary changes, but the subsequent adaptation that they bring about is due to random changes in the interactions between the species involved, and may occur even if the environment remains constant.

With a simple bookkeeping of the amount of resources that are taken from the environment, of those that are excreted as byproducts of metabolic reactions, and of the amount of byproducts from the other species that can be re-used for their own purpose, the model can cover virtually all possible ecological interactions between two species, and show, using adaptive dynamics, that these interactions evolve, going through different scenarios until reaching a final stable state. The model also provides some clues about general trends that can occur in real situations. For instance, there is a marked trend toward the emergence of mutualistic interactions, even in systems that start in competition or show antagonistic relationships. Also, many of the observed pathways have a real counterpart, because similar ones have been documented for actual species (microbial or otherwise).

The model relies upon the availability of a phenomenological model of ecological interactions \cite{garcia-algarra:2014,stucchi:2020} that is capable of describing mutualistic, competitive, and antagonistic interactions with a simple tuning of the parameters---in a way that generalized Lotka-Volterra models are not capable of. This general logistic model of population dynamics is one of the simplest models capable of exhibiting evolutionary transitions between different ecological regimes. The clue in devising our present evolutionary model has been linking those parameters to microscopic interactions between the species involved---which we have chosen as microbes for the sake of simplicity.

There are at least two ways in which this work can be extended. One is making more detailed and realistic assumptions on the microscopic interactions that occur between the two species and see whether the trends observed in this toy model are or are not confirmed. Obviously, introducing further details will impose constraints that the present model is currently free of. These constraints will bias the distribution of scenarios we observe in different ways, and will do so differently for different species---for which the details can vary from instance to instance. This will provide interesting information about the connection between microscopic mechanisms and ecological transitions.

The other way to extend this work is to consider more than two species. We can only imagine the richness of ecological scenarios that such an extension will reveal, even with a mechanism as simple as the straightforward bookkeeping of resources we have implemented. The computational complexities of this extension are evident by just looking at the involved calculations that only two species has led to. This is one of the reasons why we have decided to postpone such a study for future research---the other one being that just two species are enough to provide the proof of concept we aimed at with this work.

\begin{acknowledgments}
This research has been funded by the Spanish Ministerio de Ciencia, Innovaci\'on y Universidades-FEDER funds of the European Union support, under projects BASIC (PGC2018-098186-B-I00, J.A.C., and  PGC2018-093854-B-I00, J.G. and J.M.P.), and EVA (CGL2016-77377-R, J.M.I).
\end{acknowledgments}

\appendix

\section{Generalised logistic model of population dynamics}
\label{box:logistic}
The idea behind the population model of \cite{stucchi:2020} is to extend Velhurst's logistic equations of populations
\begin{equation}
\dot{N}_i=N_i\left(\bar r_i-\bar a_i N_i\right), \quad i=1,\dots p,
\label{eq:logistic}
\end{equation}
by making the parameters $\bar r_i$ and $\bar a_i$ to depend on the interactions with the environment as well as the population sizes of all species in the community as
\begin{equation}
\bar r_i=r_i+\sum_{k=1}^pb_{ik}N_k, \qquad
\bar a_i=a_i+c_i\sum_{k=1}^pb_{ik}N_k.
\label{eq:riai}
\end{equation}
Here $r_i$ is the intrinsic growth rate of species $i$, $b_{ik}$ is the rate of benefit (if positive) or hindrance (if negative) on species $i$ due to the interaction with species $k$, and $p$ is the total number of species in the ecosystem. The coefficients $a_i$ measure intraspecific competitions (hence $a_i>0$) due to a limitation of the environmental resources. As a matter of fact, in the standard Velhurst's model $a_i^{-1}$ directly measures the carrying capacity of the environment. As of $c_i$, the effect of these coefficients is better seen if we rewrite Eq.~\eqref{eq:logistic} and Eq.~\eqref{eq:riai} as
\begin{equation}
\dot{N}_i=N_i\left(r_i-a_i N_i\right)+N_i\left(1-c_i N_i\right)\sum_{k=1}^pb_{ik}N_k.
\label{eq:generalised}
\end{equation}
We can clearly see in this expression that choosing $c_i>0$ induces a saturation on the interaction of the community with the focal species.


For only two species the stationary solutions can be obtained as:
\begin{equation}
\begin{split}
    N^{\ast}_1=\frac{-a_1a_2 + b_{12}b_{21} + b_{21}c_2r_1 - b_{12}c_1r_2 \pm \sqrt{\Delta}}{2b_{21}(b_{12}c_1 + a_1c_2)},\\
    N^{\ast}_2=\frac{-a_{1}a_{2} + b_{12}b_{21} - b_{21}c_{2}r_{1} + b_{12}c_{1}r_{2} \pm \sqrt{\Delta}}{2b_{12}(b_{21}c_{2}+ a_{2}c_{1})},
\end{split}
\label{eq:N1N2}
\end{equation}
where
\begin{equation}
\begin{split}
    \Delta=&\, 4b_{21}(b_{12}c_1 + a_1c_2)(a_2r_1 + b_{12}r_2)\\
    &+[a_1a_2-b_{21}(b_{12}+c_2r_1)+b_{12}c_1r_2]^2.
\end{split}
\end{equation}

According to \cite{stucchi:2020}, the linear stability of the finite stationary solution for two species ($N^{\ast}_1,N^{\ast}_2$) can be analyzed from the Jacobian matrix
\begin{equation}
\label{eq:Jacob_x1x2}
J(N^{\ast}_1,N^{\ast}_2) = \left(
\begin{array}{cc}
-r_{1}-b_{12}N^{\ast}_2 & b_{12}N^{\ast}_1\left[1-c_{1}N^{\ast}_1\right] \\
b_{21}N^{\ast}_2\left[1-c_{2}N^{\ast}_2\right] & -r_{2}-b_{21}N^{\ast}_1
\end{array}
\right)
\end{equation}

\section{Adaptive dynamics equations for the two-species ecosystem}
\label{app1}

According to Eq.~\eqref{eq:2population}, the per-capita fitness of species $i$ is given by
\begin{equation}
f_i(\Omega,N_1,N_2)=r_i(\Omega)-a_iN_1+b_{ij}(\Omega)(1-c_iN_i)N_j,
\end{equation}
with $r_i(\Omega)$ and $b_{ij}(\Omega)$ given by Eq.~\eqref{eq:ripars} and Eq.~\eqref{eq:bijpars} respectively. If the parameters of the mutant are $u'_i=u_i+\delta u_i$, $w'=w+\delta w$, with $\delta u_i=\pm 1/K$ and $\delta w=0,\pm 1/K$ (see Table~\ref{tab:transitions}), then---given the linearity of $f_i$ with respect to these parameters---the per-capita fitness of the mutant in a steady state community can be written as
\begin{equation}
\bar f_i(u'_i,w',\Omega,N_1^*,N_2^*)=f_{i,u}\delta u_i-f_{i,w}\delta w,
\label{eq:fitnessmutant}
\end{equation}
where
\begin{equation}
\begin{split}
f_{i,u} &= \frac{r_{i0}(1-\gamma-\alpha_1\alpha_2)+b_{i0}\alpha_1\alpha_2(1-c_iN_i^*)N_j^*}{1-\alpha_1\alpha_2}, \\
f_{i,w} &= b_{i0}(1-c_iN_i^*)N_j^*.
\end{split}
\label{eq:fitness_f_u_w}
\end{equation}
Accordingly, the canonical equations of AD for this sort of mutants read (see \citep{dieckmann:1996})
\begin{widetext}
\begin{equation}
\begin{split}
\dot u_1&=\mu N_1^*\sum_{\delta u_1,\delta w}\delta u_1p_1(\delta u_1,\delta w)\big[\bar f_1(u_1+\delta u_1,w+\delta w,\Omega,N_1^*,N_2^*)\big]_+, \\
\dot u_2&=\mu N_2^*\sum_{\delta u_2,\delta w}\delta u_2p_2(\delta u_2,\delta w)\big[\bar f_2(u_2+\delta u_2,w+\delta w,\Omega,N_1^*,N_2^*)\big]_+, \\
\dot w&=\mu N_1^*\sum_{\delta u_1,\delta w}\delta wp_1(\delta u_1,\delta w)\big[\bar f_1(u_1+\delta u_1,w+\delta w,\Omega,N_1^*,N_2^*)\big]_+ \\
&\phantom{=}+\mu N_2^*\sum_{\delta u_2,\delta w}\delta wp_2(\delta u_2,\delta w)\big[\bar f_2(u_2+\delta u_2,w+\delta w,\Omega,N_1^*,N_2^*)\big]_+,
\end{split}
\end{equation}
\end{widetext}
where the sums run over all the corresponding mutations and $p_i(\delta u_i,\delta w)$ are their respective probabilities, according to Table~\ref{tab:transitions}, and $\mu$ is the probability of mutation per reproduction event. The function $[x]_+$ stands for $x$ if $x\ge 0$ and $0$ otherwise.

Substituting Eq.~\eqref{eq:fitnessmutant} and Eq.~\eqref{eq:fitness_f_u_w}, performing the sums, and re-scaling evolutionary time with $2K^2/\mu$ yields the set of differential equations
\begin{equation}
\begin{split}
\dot u_1&=2N_1^*\left\{[f_{1,u}-f_{1,w}]_+(u_2-w)-w[f_{1,w}-f_{1,u}]_+\right. \\
&\phantom{=}\left.+[f_{1,u}]_+(1-u_1-u_2+w)-[-f_{1,u}]_+(u_1-w)\right\}, \\
\dot u_2 &=2N_2^*\left\{[f_{2,u}-f_{2,w}]_+(u_1-w)-w[f_{2,w}-f_{2,u}]_+\right. \\
&\phantom{=}\left.+[f_{2,u}]_+(1-u_1-u_2+w)-[-f_{2,u}]_+(u_2-w)\right\}, \\
\dot w &=2N_1^*\left\{[f_{1,u}-f_{1,w}]_+(u_2-w)-w[f_{1,w}-f_{1,u}]_+\right\} \\
&\phantom{=}+2N_2^*\left\{[f_{2,u}-f_{2,w}]_+(u_1-w)-w[f_{2,w}-f_{2,u}]_+\right\},
\end{split}
\end{equation}
which, using the identity $[x]_+=(x+|x|)/2$, can be rewritten as
\begin{equation}
\begin{split}
\dot u_1 &=N_1^*\left\{f_{1,u}-f_{1,w}u_2+|f_{1,u}-f_{1,w}|(u_2-2w)\right.\\
&\phantom{=}\left.+|f_{1,u}|(1-u_2-2u_1+2w)\right\}, \\
\dot u_2 &=N_2^*\left\{f_{2,u}-f_{2,w}u_1+|f_{2,u}-f_{2,w}|(u_1-2w)\right.\\
&\phantom{=}\left.+|f_{2,u}|(1-u_1-2u_2+2w)\right\}, \\
\dot w &=N_1^*\left\{(f_{1,u}-f_{1,w})u_2+|f_{1,u}-f_{1,w}|(u_2-2w)\right\} \\
&\phantom{=}+N_2^*\left\{(f_{2,u}-f_{2,w})u_1+|f_{2,u}-f_{2,w}|(u_1-2w)\right\}.
\end{split}
\label{eq:evolut_u1_u2_w}
\end{equation}

On its side, mutants that change their parameter $\alpha_i$ have a per-capita fitness
\begin{equation}
\bar f_i(\alpha'_i,\Omega,N_1^*,N_2^*)=f_{i,\alpha}\delta\bar\alpha_i+O\left(\delta\bar\alpha_i^2\right),
\label{eq:fitnessmutant2}
\end{equation}
where $\delta\bar\alpha_i=\pm 1/K$ and
\begin{equation}
f_{i,\alpha}=\frac{b_{i0}(1-c_iN_i^*)N_j^*-r_{i0}\gamma}{1-\alpha_1\alpha_2}.
\end{equation}
Accordingly,
\begin{equation}
\begin{split}
\dot\alpha_1 &=\mu N_1^*\sum_{\delta\alpha_1}\delta\alpha_1p_1(\delta\alpha_1)
\big[\bar f_1(\alpha_1+\delta\alpha_1,\Omega,N_1^*,N_2^*)\big]_+, \\
\dot\alpha_2 &=\mu N_2^*\sum_{\delta\alpha_2}\delta\alpha_2p_2(\delta\alpha_2)
\big[\bar f_2(\alpha_2+\delta\alpha_2,\Omega,N_1^*,N_2^*)\big]_+,
\end{split}
\end{equation}
which, neglecting $O\left(\delta\bar\alpha_i^2\right)$ terms and re-scaling time as before, becomes
\begin{equation}
\begin{split}
\dot\alpha_1 &=\frac{2N_1^*}{\gamma}\left\{[f_{1,\alpha}]_+(\gamma-\alpha_1)
-[-f_{1,\alpha}]_+\alpha_1\right\} \\
&=N_1^*\left\{f_{1,\alpha}
+|f_{1,\alpha}|\frac{\gamma-2\alpha_1}{\gamma}\right\}, \\
\dot\alpha_2 &=\frac{2N_2^*}{\gamma}\left\{[f_{2,\alpha}]_+(\gamma-\alpha_2)
-[-f_{2,\alpha}]_+\alpha_2\right\} \\
&=N_2^*\left\{f_{2,\alpha}
+|f_{2,\alpha}|\frac{\gamma-2\alpha_2}{\gamma}\right\}.
\end{split}
\label{eq:evolut_alpha}
\end{equation}
\vspace*{8mm}
\ \\


\bibliography{references}

\begin{thebibliography}{44}%
\makeatletter
\providecommand \@ifxundefined [1]{%
 \@ifx{#1\undefined}
}%
\providecommand \@ifnum [1]{%
 \ifnum #1\expandafter \@firstoftwo
 \else \expandafter \@secondoftwo
 \fi
}%
\providecommand \@ifx [1]{%
 \ifx #1\expandafter \@firstoftwo
 \else \expandafter \@secondoftwo
 \fi
}%
\providecommand \natexlab [1]{#1}%
\providecommand \enquote  [1]{``#1''}%
\providecommand \bibnamefont  [1]{#1}%
\providecommand \bibfnamefont [1]{#1}%
\providecommand \citenamefont [1]{#1}%
\providecommand \href@noop [0]{\@secondoftwo}%
\providecommand \href [0]{\begingroup \@sanitize@url \@href}%
\providecommand \@href[1]{\@@startlink{#1}\@@href}%
\providecommand \@@href[1]{\endgroup#1\@@endlink}%
\providecommand \@sanitize@url [0]{\catcode `\\12\catcode `\$12\catcode
  `\&12\catcode `\#12\catcode `\^12\catcode `\_12\catcode `\%12\relax}%
\providecommand \@@startlink[1]{}%
\providecommand \@@endlink[0]{}%
\providecommand \url  [0]{\begingroup\@sanitize@url \@url }%
\providecommand \@url [1]{\endgroup\@href {#1}{\urlprefix }}%
\providecommand \urlprefix  [0]{URL }%
\providecommand \Eprint [0]{\href }%
\providecommand \doibase [0]{https://doi.org/}%
\providecommand \selectlanguage [0]{\@gobble}%
\providecommand \bibinfo  [0]{\@secondoftwo}%
\providecommand \bibfield  [0]{\@secondoftwo}%
\providecommand \translation [1]{[#1]}%
\providecommand \BibitemOpen [0]{}%
\providecommand \bibitemStop [0]{}%
\providecommand \bibitemNoStop [0]{.\EOS\space}%
\providecommand \EOS [0]{\spacefactor3000\relax}%
\providecommand \BibitemShut  [1]{\csname bibitem#1\endcsname}%
\let\auto@bib@innerbib\@empty
\bibitem [{\citenamefont {Park}\ \emph {et~al.}(2004)\citenamefont {Park},
  \citenamefont {Yun}, \citenamefont {Kim}, \citenamefont {Chun},\ and\
  \citenamefont {Ahn}}]{park:2004}%
  \BibitemOpen
  \bibfield  {author} {\bibinfo {author} {\bibfnamefont {M.}~\bibnamefont
  {Park}}, \bibinfo {author} {\bibfnamefont {S.}~\bibnamefont {Yun}}, \bibinfo
  {author} {\bibfnamefont {M.}~\bibnamefont {Kim}}, \bibinfo {author}
  {\bibfnamefont {J.}~\bibnamefont {Chun}},\ and\ \bibinfo {author}
  {\bibfnamefont {T.}~\bibnamefont {Ahn}},\ }\bibfield  {title} {\bibinfo
  {title} {{Phylogenetic characterization of Legionella-like endosymbiotic
  X-bacteria in Amoeba proteus: a proposal for `Candidatus Legionella jeonii'
  sp. nov.}},\ }\href@noop {} {\bibfield  {journal} {\bibinfo  {journal}
  {Environ. Microbiol.}\ }\textbf {\bibinfo {volume} {6}},\ \bibinfo {pages}
  {1252} (\bibinfo {year} {2004})}\BibitemShut {NoStop}%
\bibitem [{\citenamefont {Jeon}\ and\ \citenamefont {Lorch}(1967)}]{jeon:1967}%
  \BibitemOpen
  \bibfield  {author} {\bibinfo {author} {\bibfnamefont {K.~W.}\ \bibnamefont
  {Jeon}}\ and\ \bibinfo {author} {\bibfnamefont {I.~J.}\ \bibnamefont
  {Lorch}},\ }\bibfield  {title} {\bibinfo {title} {Unusual intra-cellular
  bacterial infection in large, free-living amoebae},\ }\href@noop {}
  {\bibfield  {journal} {\bibinfo  {journal} {Exp. Cell Res.}\ }\textbf
  {\bibinfo {volume} {48}},\ \bibinfo {pages} {236} (\bibinfo {year}
  {1967})}\BibitemShut {NoStop}%
\bibitem [{\citenamefont {Jeon}(1992)}]{jeon:1992}%
  \BibitemOpen
  \bibfield  {author} {\bibinfo {author} {\bibfnamefont {K.~W.}\ \bibnamefont
  {Jeon}},\ }\bibfield  {title} {\bibinfo {title} {Macromolecules involved in
  the amoeba-bacteria symbiosis},\ }\href@noop {} {\bibfield  {journal}
  {\bibinfo  {journal} {J. Protozool.}\ }\textbf {\bibinfo {volume} {39}},\
  \bibinfo {pages} {199} (\bibinfo {year} {1992})}\BibitemShut {NoStop}%
\bibitem [{\citenamefont {Jeon}(1995)}]{jeon:1995}%
  \BibitemOpen
  \bibfield  {author} {\bibinfo {author} {\bibfnamefont {K.~W.}\ \bibnamefont
  {Jeon}},\ }\bibfield  {title} {\bibinfo {title} {Bacterial endosymbiosis in
  amoebae},\ }\href {https://doi.org/10.1016/S0962-8924(00)88966-7} {\bibfield
  {journal} {\bibinfo  {journal} {Trends Cell. Biol.}\ }\textbf {\bibinfo
  {volume} {5}},\ \bibinfo {pages} {137} (\bibinfo {year} {1995})}\BibitemShut
  {NoStop}%
\bibitem [{\citenamefont {Machado}\ \emph {et~al.}(2001)\citenamefont
  {Machado}, \citenamefont {Jousselin}, \citenamefont {Kjellberg},
  \citenamefont {Compton},\ and\ \citenamefont {Herre}}]{machado:2001}%
  \BibitemOpen
  \bibfield  {author} {\bibinfo {author} {\bibfnamefont {C.~A.}\ \bibnamefont
  {Machado}}, \bibinfo {author} {\bibfnamefont {E.}~\bibnamefont {Jousselin}},
  \bibinfo {author} {\bibfnamefont {F.}~\bibnamefont {Kjellberg}}, \bibinfo
  {author} {\bibfnamefont {S.~G.}\ \bibnamefont {Compton}},\ and\ \bibinfo
  {author} {\bibfnamefont {E.~A.}\ \bibnamefont {Herre}},\ }\bibfield  {title}
  {\bibinfo {title} {Phylogenetic relationships, historical biogeography and
  character evolution of fig-pollinating wasps},\ }\href@noop {} {\bibfield
  {journal} {\bibinfo  {journal} {Proc. R. Soc. B}\ }\textbf {\bibinfo {volume}
  {268}},\ \bibinfo {pages} {685} (\bibinfo {year} {2001})}\BibitemShut
  {NoStop}%
\bibitem [{\citenamefont {Shingleton}\ and\ \citenamefont
  {Stern}(2002)}]{shingleton:2002}%
  \BibitemOpen
  \bibfield  {author} {\bibinfo {author} {\bibfnamefont {A.~W.}\ \bibnamefont
  {Shingleton}}\ and\ \bibinfo {author} {\bibfnamefont {D.~L.}\ \bibnamefont
  {Stern}},\ }\bibfield  {title} {\bibinfo {title} {Molecular phylogenetic
  evidence for multiple gains or losses of ant mutualism within the aphid genus
  \textit{Chaitophorus}},\ }\href@noop {} {\bibfield  {journal} {\bibinfo
  {journal} {Mol. Phylogenet. Evol.}\ }\textbf {\bibinfo {volume} {26}},\
  \bibinfo {pages} {26} (\bibinfo {year} {2002})}\BibitemShut {NoStop}%
\bibitem [{\citenamefont {Sakata}(1994)}]{sakata:1994}%
  \BibitemOpen
  \bibfield  {author} {\bibinfo {author} {\bibfnamefont {H.}~\bibnamefont
  {Sakata}},\ }\bibfield  {title} {\bibinfo {title} {How an ant decides to prey
  on or to attend aphids},\ }\href@noop {} {\bibfield  {journal} {\bibinfo
  {journal} {Res. Popul. Ecol.}\ }\textbf {\bibinfo {volume} {36}},\ \bibinfo
  {pages} {45} (\bibinfo {year} {1994})}\BibitemShut {NoStop}%
\bibitem [{\citenamefont {Stadler}\ and\ \citenamefont
  {Dixon}(2005)}]{stadler:2005}%
  \BibitemOpen
  \bibfield  {author} {\bibinfo {author} {\bibfnamefont {B.}~\bibnamefont
  {Stadler}}\ and\ \bibinfo {author} {\bibfnamefont {A.~F.~G.}\ \bibnamefont
  {Dixon}},\ }\bibfield  {title} {\bibinfo {title} {Ecology and evolution of
  aphid-ant interactions},\ }\href@noop {} {\bibfield  {journal} {\bibinfo
  {journal} {Annu. Rev. Ecol. Evol. Syst.}\ }\textbf {\bibinfo {volume} {36}},\
  \bibinfo {pages} {345} (\bibinfo {year} {2005})}\BibitemShut {NoStop}%
\bibitem [{\citenamefont {Offenberg}(2001)}]{offenberg:2001}%
  \BibitemOpen
  \bibfield  {author} {\bibinfo {author} {\bibfnamefont {J.}~\bibnamefont
  {Offenberg}},\ }\bibfield  {title} {\bibinfo {title} {{Balancing between
  mutualism and exploitation: the symbiotic interaction between \textit{Lasius}
  ants and aphids}},\ }\href@noop {} {\bibfield  {journal} {\bibinfo  {journal}
  {Behav. Ecol. Sociobiol.}\ }\textbf {\bibinfo {volume} {49}},\ \bibinfo
  {pages} {304} (\bibinfo {year} {2001})}\BibitemShut {NoStop}%
\bibitem [{\citenamefont {Sachs}\ and\ \citenamefont
  {Simms}(2006)}]{sachs:2006}%
  \BibitemOpen
  \bibfield  {author} {\bibinfo {author} {\bibfnamefont {J.~L.}\ \bibnamefont
  {Sachs}}\ and\ \bibinfo {author} {\bibfnamefont {E.~L.}\ \bibnamefont
  {Simms}},\ }\bibfield  {title} {\bibinfo {title} {Pathways to mutualism
  breakdown},\ }\href@noop {} {\bibfield  {journal} {\bibinfo  {journal}
  {Trends Ecol. Evol.}\ }\textbf {\bibinfo {volume} {21}},\ \bibinfo {pages}
  {585} (\bibinfo {year} {2006})}\BibitemShut {NoStop}%
\bibitem [{\citenamefont {Dieckmann}\ and\ \citenamefont
  {Law}(1996)}]{dieckmann:1996}%
  \BibitemOpen
  \bibfield  {author} {\bibinfo {author} {\bibfnamefont {U.}~\bibnamefont
  {Dieckmann}}\ and\ \bibinfo {author} {\bibfnamefont {R.}~\bibnamefont
  {Law}},\ }\bibfield  {title} {\bibinfo {title} {The dynamical theory of
  coevolution: a derivation from stochastic ecological processes},\ }\href
  {https://doi.org/10.1007/BF02409751} {\bibfield  {journal} {\bibinfo
  {journal} {J. Math. Biol.}\ }\textbf {\bibinfo {volume} {34}},\ \bibinfo
  {pages} {579} (\bibinfo {year} {1996})}\BibitemShut {NoStop}%
\bibitem [{\citenamefont {Dercole}\ and\ \citenamefont
  {Rinaldi}(2008)}]{dercole:2008}%
  \BibitemOpen
  \bibfield  {author} {\bibinfo {author} {\bibfnamefont {F.}~\bibnamefont
  {Dercole}}\ and\ \bibinfo {author} {\bibfnamefont {S.}~\bibnamefont
  {Rinaldi}},\ }\href@noop {} {\emph {\bibinfo {title} {Analysis of
  Evolutionary Processes: The Adaptive Dynamics Approach and Its
  Applications}}}\ (\bibinfo  {publisher} {Princeton University Press},\
  \bibinfo {address} {Princeton, New Jersey},\ \bibinfo {year}
  {2008})\BibitemShut {NoStop}%
\bibitem [{\citenamefont {Doebeli}(2011)}]{doebeli:2011}%
  \BibitemOpen
  \bibfield  {author} {\bibinfo {author} {\bibfnamefont {M.}~\bibnamefont
  {Doebeli}},\ }\href@noop {} {\emph {\bibinfo {title} {Adaptive
  Diversification}}}\ (\bibinfo  {publisher} {Princeton University Press},\
  \bibinfo {address} {Princeton, New Jersey},\ \bibinfo {year}
  {2011})\BibitemShut {NoStop}%
\bibitem [{\citenamefont {Lotka}(1925)}]{lotka:1925}%
  \BibitemOpen
  \bibfield  {author} {\bibinfo {author} {\bibfnamefont {A.~J.}\ \bibnamefont
  {Lotka}},\ }\href@noop {} {\emph {\bibinfo {title} {{Elements of Physical
  Biology}}}}\ (\bibinfo  {publisher} {Williams and Wilkins Company},\ \bibinfo
  {address} {Baltimore},\ \bibinfo {year} {1925})\BibitemShut {NoStop}%
\bibitem [{\citenamefont {Volterra}(1926)}]{volterra:1926}%
  \BibitemOpen
  \bibfield  {author} {\bibinfo {author} {\bibfnamefont {V.}~\bibnamefont
  {Volterra}},\ }\bibfield  {title} {\bibinfo {title} {Fluctuations in the
  abundance of a species considered mathematically},\ }\href@noop {} {\bibfield
   {journal} {\bibinfo  {journal} {Nature}\ }\textbf {\bibinfo {volume}
  {118}},\ \bibinfo {pages} {558} (\bibinfo {year} {1926})}\BibitemShut
  {NoStop}%
\bibitem [{\citenamefont {Volterra}(1928)}]{volterra:1928}%
  \BibitemOpen
  \bibfield  {author} {\bibinfo {author} {\bibfnamefont {V.}~\bibnamefont
  {Volterra}},\ }\bibfield  {title} {\bibinfo {title} {{Variations and
  Fluctuations of the Number of Individuals in Animal Species Living
  Together}},\ }\href@noop {} {\bibfield  {journal} {\bibinfo  {journal} {ICES
  J. Mar. Sci.}\ }\textbf {\bibinfo {volume} {3}},\ \bibinfo {pages} {3}
  (\bibinfo {year} {1928})}\BibitemShut {NoStop}%
\bibitem [{\citenamefont {May}(1981)}]{may:1981}%
  \BibitemOpen
  \bibfield  {author} {\bibinfo {author} {\bibfnamefont {R.~M.}\ \bibnamefont
  {May}},\ }\bibfield  {title} {\bibinfo {title} {Models for two interacting
  populations},\ }in\ \href@noop {} {\emph {\bibinfo {booktitle} {Theoretical
  Ecology: Principles and Applications}}},\ \bibinfo {editor} {edited by\
  \bibinfo {editor} {\bibfnamefont {R.~M.}\ \bibnamefont {May}}\ and\ \bibinfo
  {editor} {\bibfnamefont {A.~R.}\ \bibnamefont {McLean}}}\ (\bibinfo
  {publisher} {Oxford Univesity Press},\ \bibinfo {address} {Oxford, UK},\
  \bibinfo {year} {1981})\ pp.\ \bibinfo {pages} {78--104}\BibitemShut
  {NoStop}%
\bibitem [{\citenamefont {Wright}(1989)}]{wright:1989}%
  \BibitemOpen
  \bibfield  {author} {\bibinfo {author} {\bibfnamefont {D.~H.}\ \bibnamefont
  {Wright}},\ }\bibfield  {title} {\bibinfo {title} {A simple, stable model of
  mutualism incorporating handling time},\ }\href@noop {} {\bibfield  {journal}
  {\bibinfo  {journal} {Am. Nat.}\ }\textbf {\bibinfo {volume} {134}},\
  \bibinfo {pages} {664} (\bibinfo {year} {1989})}\BibitemShut {NoStop}%
\bibitem [{\citenamefont {Garc{\'{\i}}a-Algarra}\ \emph
  {et~al.}(2014)\citenamefont {Garc{\'{\i}}a-Algarra}, \citenamefont {Galeano},
  \citenamefont {Pastor}, \citenamefont {Iriondo},\ and\ \citenamefont
  {Ramasco}}]{garcia-algarra:2014}%
  \BibitemOpen
  \bibfield  {author} {\bibinfo {author} {\bibfnamefont {J.}~\bibnamefont
  {Garc{\'{\i}}a-Algarra}}, \bibinfo {author} {\bibfnamefont {J.}~\bibnamefont
  {Galeano}}, \bibinfo {author} {\bibfnamefont {J.~M.}\ \bibnamefont {Pastor}},
  \bibinfo {author} {\bibfnamefont {J.~M.}\ \bibnamefont {Iriondo}},\ and\
  \bibinfo {author} {\bibfnamefont {J.~J.}\ \bibnamefont {Ramasco}},\
  }\bibfield  {title} {\bibinfo {title} {Rethinking the logistic approach for
  population dynamics of mutualistic interactions},\ }\href@noop {} {\bibfield
  {journal} {\bibinfo  {journal} {J. Theor. Biol.}\ }\textbf {\bibinfo {volume}
  {363}},\ \bibinfo {pages} {332} (\bibinfo {year} {2014})}\BibitemShut
  {NoStop}%
\bibitem [{\citenamefont {Stucchi}\ \emph {et~al.}(2020)\citenamefont
  {Stucchi}, \citenamefont {Pastor}, \citenamefont {García-Algarra},\ and\
  \citenamefont {Galeano}}]{stucchi:2020}%
  \BibitemOpen
  \bibfield  {author} {\bibinfo {author} {\bibfnamefont {L.}~\bibnamefont
  {Stucchi}}, \bibinfo {author} {\bibfnamefont {J.~M.}\ \bibnamefont {Pastor}},
  \bibinfo {author} {\bibfnamefont {J.}~\bibnamefont {García-Algarra}},\ and\
  \bibinfo {author} {\bibfnamefont {J.}~\bibnamefont {Galeano}},\ }\bibfield
  {title} {\bibinfo {title} {A general model of population dynamics accounting
  for multiple kinds of interaction},\ }\bibfield  {journal} {\bibinfo
  {journal} {Complexity}\ }\textbf {\bibinfo {volume} {2020}},\ \href
  {https://doi.org/10.1155/2020/7961327} {10.1155/2020/7961327} (\bibinfo
  {year} {2020})\BibitemShut {NoStop}%
\bibitem [{\citenamefont {Thompson}(1988)}]{thompson:1988}%
  \BibitemOpen
  \bibfield  {author} {\bibinfo {author} {\bibfnamefont {J.~N.}\ \bibnamefont
  {Thompson}},\ }\bibfield  {title} {\bibinfo {title} {Variation in
  interspecific interactions},\ }\href@noop {} {\bibfield  {journal} {\bibinfo
  {journal} {Annu. Rev. Ecol. Syst.}\ }\textbf {\bibinfo {volume} {19}},\
  \bibinfo {pages} {65} (\bibinfo {year} {1988})}\BibitemShut {NoStop}%
\bibitem [{\citenamefont {Hoek}\ \emph {et~al.}(2016)\citenamefont {Hoek},
  \citenamefont {Axelrod}, \citenamefont {Biancalani}, \citenamefont {Yurtsev},
  \citenamefont {Liu},\ and\ \citenamefont {Gore}}]{hoek:2016}%
  \BibitemOpen
  \bibfield  {author} {\bibinfo {author} {\bibfnamefont {T.~A.}\ \bibnamefont
  {Hoek}}, \bibinfo {author} {\bibfnamefont {K.}~\bibnamefont {Axelrod}},
  \bibinfo {author} {\bibfnamefont {T.}~\bibnamefont {Biancalani}}, \bibinfo
  {author} {\bibfnamefont {E.~A.}\ \bibnamefont {Yurtsev}}, \bibinfo {author}
  {\bibfnamefont {J.}~\bibnamefont {Liu}},\ and\ \bibinfo {author}
  {\bibfnamefont {J.}~\bibnamefont {Gore}},\ }\bibfield  {title} {\bibinfo
  {title} {Resource availability modulates the cooperative and competitive
  nature of a microbial cross-feeding mutualism},\ }\href@noop {} {\bibfield
  {journal} {\bibinfo  {journal} {PLOS Biology}\ }\textbf {\bibinfo {volume}
  {15}},\ \bibinfo {pages} {e1002606} (\bibinfo {year} {2016})}\BibitemShut
  {NoStop}%
\bibitem [{\citenamefont {Tilman}(2004)}]{tilman:2004}%
  \BibitemOpen
  \bibfield  {author} {\bibinfo {author} {\bibfnamefont {D.}~\bibnamefont
  {Tilman}},\ }\bibfield  {title} {\bibinfo {title} {Niche tradeoffs,
  neutrality, and community structure: A stochastic theory of resource
  competition, invasion, and community assembly},\ }\href@noop {} {\bibfield
  {journal} {\bibinfo  {journal} {Proc. Natl. Acad. Sci. (USA)}\ }\textbf
  {\bibinfo {volume} {101}},\ \bibinfo {pages} {10854} (\bibinfo {year}
  {2004})}\BibitemShut {NoStop}%
\bibitem [{\citenamefont {BioRender}(2019)}]{Bio}%
  \BibitemOpen
  \bibfield  {author} {\bibinfo {author} {\bibnamefont {BioRender}},\
  }\href@noop {} {} (\bibinfo {year} {2019}),\ \bibinfo {note}
  {\url{https://biorender.com}, last accessed on 2019-07-10}\BibitemShut
  {NoStop}%
\bibitem [{\citenamefont {Schwartz}\ and\ \citenamefont
  {Hoeksema}(1998)}]{schwartz:1998}%
  \BibitemOpen
  \bibfield  {author} {\bibinfo {author} {\bibfnamefont {M.~W.}\ \bibnamefont
  {Schwartz}}\ and\ \bibinfo {author} {\bibfnamefont {J.~D.}\ \bibnamefont
  {Hoeksema}},\ }\bibfield  {title} {\bibinfo {title} {Specialization and
  resource trade: biological markets as a model of mutualisms},\ }\href@noop {}
  {\bibfield  {journal} {\bibinfo  {journal} {Ecology}\ }\textbf {\bibinfo
  {volume} {79}} (\bibinfo {year} {1998})}\BibitemShut {NoStop}%
\bibitem [{sup()}]{suppmat}%
  \BibitemOpen
  \href@noop {} {}\bibinfo {note} {See Supplemental Material at [URL to be
  inserted by publisher] for details about the derivation of the model as well
  as supplementary figures.}\BibitemShut {Stop}%
\bibitem [{\citenamefont {Turner}\ and\ \citenamefont
  {Chao}(1999)}]{turner:1999}%
  \BibitemOpen
  \bibfield  {author} {\bibinfo {author} {\bibfnamefont {P.~E.}\ \bibnamefont
  {Turner}}\ and\ \bibinfo {author} {\bibfnamefont {L.}~\bibnamefont {Chao}},\
  }\bibfield  {title} {\bibinfo {title} {{Prisoner's dilemma in an RNA
  virus}},\ }\href {https://doi.org/10.1038/18913} {\bibfield  {journal}
  {\bibinfo  {journal} {Nature}\ }\textbf {\bibinfo {volume} {398}},\ \bibinfo
  {pages} {441} (\bibinfo {year} {1999})}\BibitemShut {NoStop}%
\bibitem [{\citenamefont {Harcombe}\ \emph {et~al.}(2018)\citenamefont
  {Harcombe}, \citenamefont {Chac{\'o}n}, \citenamefont {Adamowicz},
  \citenamefont {m.~Chubiz},\ and\ \citenamefont {Marx}}]{harcombe:2018}%
  \BibitemOpen
  \bibfield  {author} {\bibinfo {author} {\bibfnamefont {W.~R.}\ \bibnamefont
  {Harcombe}}, \bibinfo {author} {\bibfnamefont {J.~M.}\ \bibnamefont
  {Chac{\'o}n}}, \bibinfo {author} {\bibfnamefont {E.~M.}\ \bibnamefont
  {Adamowicz}}, \bibinfo {author} {\bibfnamefont {L.}~\bibnamefont
  {m.~Chubiz}},\ and\ \bibinfo {author} {\bibfnamefont {C.~J.}\ \bibnamefont
  {Marx}},\ }\bibfield  {title} {\bibinfo {title} {Evolution of bidirectional
  costly mutualism from byproduct consumption},\ }\href@noop {} {\bibfield
  {journal} {\bibinfo  {journal} {PNAS}\ }\textbf {\bibinfo {volume} {115}},\
  \bibinfo {pages} {12000} (\bibinfo {year} {2018})}\BibitemShut {NoStop}%
\bibitem [{\citenamefont {Nobre}\ \emph {et~al.}(2010)\citenamefont {Nobre},
  \citenamefont {Rouland-Lefevre},\ and\ \citenamefont {Aanen}}]{nobre:2010}%
  \BibitemOpen
  \bibfield  {author} {\bibinfo {author} {\bibfnamefont {T.}~\bibnamefont
  {Nobre}}, \bibinfo {author} {\bibfnamefont {C.}~\bibnamefont
  {Rouland-Lefevre}},\ and\ \bibinfo {author} {\bibfnamefont {D.~K.}\
  \bibnamefont {Aanen}},\ }\bibinfo {title} {Biology of termites: a modern
  synthesis}\ (\bibinfo  {publisher} {Springer},\ \bibinfo {address}
  {Dordrecht},\ \bibinfo {year} {2010})\ Chap.\ \bibinfo {chapter} {Comparative
  Biology of Fungus Cultivation in Termites and Ants}, pp.\ \bibinfo {pages}
  {193--210}\BibitemShut {NoStop}%
\bibitem [{\citenamefont {Aanen}\ \emph {et~al.}(2002)\citenamefont {Aanen},
  \citenamefont {Eggleton}, \citenamefont {Rouland-Lefevre}, \citenamefont
  {Guldberg-Fr{\o}slev}, \citenamefont {Rosendahl},\ and\ \citenamefont
  {Boomsma}}]{aanen:2002}%
  \BibitemOpen
  \bibfield  {author} {\bibinfo {author} {\bibfnamefont {D.~K.}\ \bibnamefont
  {Aanen}}, \bibinfo {author} {\bibfnamefont {P.}~\bibnamefont {Eggleton}},
  \bibinfo {author} {\bibfnamefont {C.}~\bibnamefont {Rouland-Lefevre}},
  \bibinfo {author} {\bibfnamefont {T.}~\bibnamefont {Guldberg-Fr{\o}slev}},
  \bibinfo {author} {\bibfnamefont {S.}~\bibnamefont {Rosendahl}},\ and\
  \bibinfo {author} {\bibfnamefont {J.~J.}\ \bibnamefont {Boomsma}},\
  }\bibfield  {title} {\bibinfo {title} {The evolution of fungus-growing
  termites and their mutualistic fungal symbionts},\ }\href@noop {} {\bibfield
  {journal} {\bibinfo  {journal} {PNAS}\ }\textbf {\bibinfo {volume} {99}},\
  \bibinfo {pages} {14887} (\bibinfo {year} {2002})}\BibitemShut {NoStop}%
\bibitem [{\citenamefont {Margulis}\ and\ \citenamefont
  {Sagan}(2002)}]{margulis:2002b}%
  \BibitemOpen
  \bibfield  {author} {\bibinfo {author} {\bibfnamefont {L.}~\bibnamefont
  {Margulis}}\ and\ \bibinfo {author} {\bibfnamefont {D.}~\bibnamefont
  {Sagan}},\ }\href@noop {} {\emph {\bibinfo {title} {Acquiring Genomes: A
  Theory of the Origin of Species}}}\ (\bibinfo  {publisher} {Basic Books},\
  \bibinfo {address} {New York, NY},\ \bibinfo {year} {2002})\BibitemShut
  {NoStop}%
\bibitem [{\citenamefont {Lutzoni}\ \emph {et~al.}(2001)\citenamefont
  {Lutzoni}, \citenamefont {Pagel},\ and\ \citenamefont {Reeb}}]{lutzoni:2001}%
  \BibitemOpen
  \bibfield  {author} {\bibinfo {author} {\bibfnamefont {F.}~\bibnamefont
  {Lutzoni}}, \bibinfo {author} {\bibfnamefont {M.}~\bibnamefont {Pagel}},\
  and\ \bibinfo {author} {\bibfnamefont {V.}~\bibnamefont {Reeb}},\ }\bibfield
  {title} {\bibinfo {title} {Major fungal lineages are derived from lichen
  symbiotic ancestors},\ }\href@noop {} {\bibfield  {journal} {\bibinfo
  {journal} {Nature}\ }\textbf {\bibinfo {volume} {411}},\ \bibinfo {pages}
  {937} (\bibinfo {year} {2001})}\BibitemShut {NoStop}%
\bibitem [{\citenamefont {Divakar}\ \emph {et~al.}(2015)\citenamefont
  {Divakar}, \citenamefont {Crespo}, \citenamefont {Wedin}, \citenamefont
  {Leavitt}, \citenamefont {Hawksworth}, \citenamefont {Myllys}, \citenamefont
  {McCune}, \citenamefont {Randlane}, \citenamefont {Bjerke}, \citenamefont
  {Ohmura}, \citenamefont {Schmitt}, \citenamefont {Boluda}, \citenamefont
  {Alors}, \citenamefont {Roca-Valiente}, \citenamefont {Del-Prado},
  \citenamefont {Ruibal}, \citenamefont {Buaruang}, \citenamefont
  {N{\'u\~{n}}ez-Zapata}, \citenamefont {de~Paz}, \citenamefont {Rico},
  \citenamefont {Molina}, \citenamefont {Elix}, \citenamefont {Esslinger},
  \citenamefont {Tronstad}, \citenamefont {Lindgren}, \citenamefont {Ertz},
  \citenamefont {Gueidan}, \citenamefont {Saag}, \citenamefont {Mark},
  \citenamefont {Singh}, \citenamefont {Grande}, \citenamefont {Parnmen},
  \citenamefont {Beck}, \citenamefont {Benatti}, \citenamefont {Blanchon},
  \citenamefont {Candan}, \citenamefont {Clerc}, \citenamefont {Goward},
  \citenamefont {Grube}, \citenamefont {Hodkinson}, \citenamefont {Hur},
  \citenamefont {Kantvilas}, \citenamefont {Kirika}, \citenamefont {Lendemer},
  \citenamefont {Mattsson}, \citenamefont {Messuti}, \citenamefont
  {Miadlikowska}, \citenamefont {Nelsen}, \citenamefont {Ohlson}, \citenamefont
  {P{\'e}rez-Ortega}, \citenamefont {Saag}, \citenamefont {Sipman},
  \citenamefont {Sohrabi}, \citenamefont {Thell}, \citenamefont {Thor},
  \citenamefont {Truong}, \citenamefont {Yahr}, \citenamefont {Upreti},
  \citenamefont {Cubas},\ and\ \citenamefont {Lumbsch}}]{divakar:2015}%
  \BibitemOpen
  \bibfield  {author} {\bibinfo {author} {\bibfnamefont {P.~K.}\ \bibnamefont
  {Divakar}}, \bibinfo {author} {\bibfnamefont {A.}~\bibnamefont {Crespo}},
  \bibinfo {author} {\bibfnamefont {M.}~\bibnamefont {Wedin}}, \bibinfo
  {author} {\bibfnamefont {S.~D.}\ \bibnamefont {Leavitt}}, \bibinfo {author}
  {\bibfnamefont {D.~L.}\ \bibnamefont {Hawksworth}}, \bibinfo {author}
  {\bibfnamefont {L.}~\bibnamefont {Myllys}}, \bibinfo {author} {\bibfnamefont
  {B.}~\bibnamefont {McCune}}, \bibinfo {author} {\bibfnamefont
  {T.}~\bibnamefont {Randlane}}, \bibinfo {author} {\bibfnamefont {J.~W.}\
  \bibnamefont {Bjerke}}, \bibinfo {author} {\bibfnamefont {Y.}~\bibnamefont
  {Ohmura}}, \bibinfo {author} {\bibfnamefont {I.}~\bibnamefont {Schmitt}},
  \bibinfo {author} {\bibfnamefont {C.~G.}\ \bibnamefont {Boluda}}, \bibinfo
  {author} {\bibfnamefont {D.}~\bibnamefont {Alors}}, \bibinfo {author}
  {\bibfnamefont {B.}~\bibnamefont {Roca-Valiente}}, \bibinfo {author}
  {\bibfnamefont {R.}~\bibnamefont {Del-Prado}}, \bibinfo {author}
  {\bibfnamefont {C.}~\bibnamefont {Ruibal}}, \bibinfo {author} {\bibfnamefont
  {K.}~\bibnamefont {Buaruang}}, \bibinfo {author} {\bibfnamefont
  {J.}~\bibnamefont {N{\'u\~{n}}ez-Zapata}}, \bibinfo {author} {\bibfnamefont
  {G.~A.}\ \bibnamefont {de~Paz}}, \bibinfo {author} {\bibfnamefont {V.~J.}\
  \bibnamefont {Rico}}, \bibinfo {author} {\bibfnamefont {M.~C.}\ \bibnamefont
  {Molina}}, \bibinfo {author} {\bibfnamefont {J.~A.}\ \bibnamefont {Elix}},
  \bibinfo {author} {\bibfnamefont {T.~L.}\ \bibnamefont {Esslinger}}, \bibinfo
  {author} {\bibfnamefont {I.~K.~K.}\ \bibnamefont {Tronstad}}, \bibinfo
  {author} {\bibfnamefont {H.}~\bibnamefont {Lindgren}}, \bibinfo {author}
  {\bibfnamefont {D.}~\bibnamefont {Ertz}}, \bibinfo {author} {\bibfnamefont
  {C.}~\bibnamefont {Gueidan}}, \bibinfo {author} {\bibfnamefont
  {L.}~\bibnamefont {Saag}}, \bibinfo {author} {\bibfnamefont {K.}~\bibnamefont
  {Mark}}, \bibinfo {author} {\bibfnamefont {G.}~\bibnamefont {Singh}},
  \bibinfo {author} {\bibfnamefont {F.~D.}\ \bibnamefont {Grande}}, \bibinfo
  {author} {\bibfnamefont {S.}~\bibnamefont {Parnmen}}, \bibinfo {author}
  {\bibfnamefont {A.}~\bibnamefont {Beck}}, \bibinfo {author} {\bibfnamefont
  {M.~N.}\ \bibnamefont {Benatti}}, \bibinfo {author} {\bibfnamefont
  {D.}~\bibnamefont {Blanchon}}, \bibinfo {author} {\bibfnamefont
  {M.}~\bibnamefont {Candan}}, \bibinfo {author} {\bibfnamefont
  {P.}~\bibnamefont {Clerc}}, \bibinfo {author} {\bibfnamefont
  {T.}~\bibnamefont {Goward}}, \bibinfo {author} {\bibfnamefont
  {M.}~\bibnamefont {Grube}}, \bibinfo {author} {\bibfnamefont {B.~P.}\
  \bibnamefont {Hodkinson}}, \bibinfo {author} {\bibfnamefont {J.-S.}\
  \bibnamefont {Hur}}, \bibinfo {author} {\bibfnamefont {G.}~\bibnamefont
  {Kantvilas}}, \bibinfo {author} {\bibfnamefont {P.~M.}\ \bibnamefont
  {Kirika}}, \bibinfo {author} {\bibfnamefont {J.}~\bibnamefont {Lendemer}},
  \bibinfo {author} {\bibfnamefont {J.-E.}\ \bibnamefont {Mattsson}}, \bibinfo
  {author} {\bibfnamefont {M.~I.}\ \bibnamefont {Messuti}}, \bibinfo {author}
  {\bibfnamefont {J.}~\bibnamefont {Miadlikowska}}, \bibinfo {author}
  {\bibfnamefont {M.}~\bibnamefont {Nelsen}}, \bibinfo {author} {\bibfnamefont
  {J.~I.}\ \bibnamefont {Ohlson}}, \bibinfo {author} {\bibfnamefont
  {S.}~\bibnamefont {P{\'e}rez-Ortega}}, \bibinfo {author} {\bibfnamefont
  {A.}~\bibnamefont {Saag}}, \bibinfo {author} {\bibfnamefont {H.~J.~M.}\
  \bibnamefont {Sipman}}, \bibinfo {author} {\bibfnamefont {M.}~\bibnamefont
  {Sohrabi}}, \bibinfo {author} {\bibfnamefont {A.}~\bibnamefont {Thell}},
  \bibinfo {author} {\bibfnamefont {G.}~\bibnamefont {Thor}}, \bibinfo {author}
  {\bibfnamefont {C.}~\bibnamefont {Truong}}, \bibinfo {author} {\bibfnamefont
  {R.}~\bibnamefont {Yahr}}, \bibinfo {author} {\bibfnamefont {D.~K.}\
  \bibnamefont {Upreti}}, \bibinfo {author} {\bibfnamefont {P.}~\bibnamefont
  {Cubas}},\ and\ \bibinfo {author} {\bibfnamefont {H.~T.}\ \bibnamefont
  {Lumbsch}},\ }\bibfield  {title} {\bibinfo {title} {Evolution of complex
  symbiotic relationships in a morphologically derived family of lichen-forming
  fungi},\ }\href@noop {} {\bibfield  {journal} {\bibinfo  {journal} {New
  Phytol.}\ }\textbf {\bibinfo {volume} {208}},\ \bibinfo {pages} {1217}
  (\bibinfo {year} {2015})}\BibitemShut {NoStop}%
\bibitem [{\citenamefont {Oppen}\ \emph {et~al.}(2005)\citenamefont {Oppen},
  \citenamefont {Mieog}, \citenamefont {S\'anchez},\ and\ \citenamefont
  {Fabricius}}]{vanoppen:2005}%
  \BibitemOpen
  \bibfield  {author} {\bibinfo {author} {\bibfnamefont {M.~J. H.~V.}\
  \bibnamefont {Oppen}}, \bibinfo {author} {\bibfnamefont {J.~. C.~.}\
  \bibnamefont {Mieog}}, \bibinfo {author} {\bibfnamefont {C.~A.}\ \bibnamefont
  {S\'anchez}},\ and\ \bibinfo {author} {\bibfnamefont {K.~E.}\ \bibnamefont
  {Fabricius}},\ }\bibfield  {title} {\bibinfo {title} {Diversity of algal
  endosymbionts (zooxanthellae) in octocorals: the roles of geography and host
  relationships},\ }\href@noop {} {\bibfield  {journal} {\bibinfo  {journal}
  {Mol Ecol}\ }\textbf {\bibinfo {volume} {14}},\ \bibinfo {pages} {2403}
  (\bibinfo {year} {2005})}\BibitemShut {NoStop}%
\bibitem [{\citenamefont {Geller}\ and\ \citenamefont
  {Walton}(2001)}]{geller:2001}%
  \BibitemOpen
  \bibfield  {author} {\bibinfo {author} {\bibfnamefont {J.~B.}\ \bibnamefont
  {Geller}}\ and\ \bibinfo {author} {\bibfnamefont {E.~D.}\ \bibnamefont
  {Walton}},\ }\bibfield  {title} {\bibinfo {title} {Breaking up and getting
  together: evolution of symbiosis and cloning by fission in sea anemones
  (genus \textit{Anthopleura})},\ }\href@noop {} {\bibfield  {journal}
  {\bibinfo  {journal} {Evolution}\ }\textbf {\bibinfo {volume} {55}},\
  \bibinfo {pages} {1781} (\bibinfo {year} {2001})}\BibitemShut {NoStop}%
\bibitem [{\citenamefont {Wilcox}(1998)}]{wilcox:1998}%
  \BibitemOpen
  \bibfield  {author} {\bibinfo {author} {\bibfnamefont {T.~P.}\ \bibnamefont
  {Wilcox}},\ }\bibfield  {title} {\bibinfo {title} {{Large-Subunit ribosomal
  RNA systematics of symbiotic dinoflagellates: morphology does not
  recapitulate phylogeny}},\ }\href@noop {} {\bibfield  {journal} {\bibinfo
  {journal} {Mol. Phylogenet. Evol.}\ }\textbf {\bibinfo {volume} {10}},\
  \bibinfo {pages} {436} (\bibinfo {year} {1998})}\BibitemShut {NoStop}%
\bibitem [{\citenamefont {Clark}\ and\ \citenamefont
  {Roger}(1995)}]{clark:1995}%
  \BibitemOpen
  \bibfield  {author} {\bibinfo {author} {\bibfnamefont {C.~G.}\ \bibnamefont
  {Clark}}\ and\ \bibinfo {author} {\bibfnamefont {A.~J.}\ \bibnamefont
  {Roger}},\ }\bibfield  {title} {\bibinfo {title} {Direct evidence for
  secondary loss of mitochondria in \textit{Entamoeba histolytica}},\
  }\href@noop {} {\bibfield  {journal} {\bibinfo  {journal} {PNAS}\ }\textbf
  {\bibinfo {volume} {92}},\ \bibinfo {pages} {6518} (\bibinfo {year}
  {1995})}\BibitemShut {NoStop}%
\bibitem [{\citenamefont {Hannaert}\ \emph {et~al.}(2003)\citenamefont
  {Hannaert}, \citenamefont {Saavedra}, \citenamefont {Duffieux}, \citenamefont
  {Szikora}, \citenamefont {Rigden}, \citenamefont {Michels},\ and\
  \citenamefont {Opperdoes}}]{hannaert:2003}%
  \BibitemOpen
  \bibfield  {author} {\bibinfo {author} {\bibfnamefont {V.}~\bibnamefont
  {Hannaert}}, \bibinfo {author} {\bibfnamefont {E.}~\bibnamefont {Saavedra}},
  \bibinfo {author} {\bibfnamefont {F.}~\bibnamefont {Duffieux}}, \bibinfo
  {author} {\bibfnamefont {J.-P.}\ \bibnamefont {Szikora}}, \bibinfo {author}
  {\bibfnamefont {D.~J.}\ \bibnamefont {Rigden}}, \bibinfo {author}
  {\bibfnamefont {P.~A.~M.}\ \bibnamefont {Michels}},\ and\ \bibinfo {author}
  {\bibfnamefont {F.~R.}\ \bibnamefont {Opperdoes}},\ }\bibfield  {title}
  {\bibinfo {title} {Plant-like traits associated with metabolism of
  \textit{Trypanosoma} parasites},\ }\href@noop {} {\bibfield  {journal}
  {\bibinfo  {journal} {PNAS}\ }\textbf {\bibinfo {volume} {100}},\ \bibinfo
  {pages} {1067} (\bibinfo {year} {2003})}\BibitemShut {NoStop}%
\bibitem [{\citenamefont {Casiraghi}\ \emph {et~al.}(2004)\citenamefont
  {Casiraghi}, \citenamefont {Bain}, \citenamefont {Guerrero}, \citenamefont
  {Martin}, \citenamefont {Pocacqua}, \citenamefont {Gardner}, \citenamefont
  {Franceschi},\ and\ \citenamefont {Bandi}}]{casiraghi:2004}%
  \BibitemOpen
  \bibfield  {author} {\bibinfo {author} {\bibfnamefont {M.}~\bibnamefont
  {Casiraghi}}, \bibinfo {author} {\bibfnamefont {O.}~\bibnamefont {Bain}},
  \bibinfo {author} {\bibfnamefont {R.}~\bibnamefont {Guerrero}}, \bibinfo
  {author} {\bibfnamefont {C.}~\bibnamefont {Martin}}, \bibinfo {author}
  {\bibfnamefont {V.}~\bibnamefont {Pocacqua}}, \bibinfo {author}
  {\bibfnamefont {S.~L.}\ \bibnamefont {Gardner}}, \bibinfo {author}
  {\bibfnamefont {A.}~\bibnamefont {Franceschi}},\ and\ \bibinfo {author}
  {\bibfnamefont {C.}~\bibnamefont {Bandi}},\ }\bibfield  {title} {\bibinfo
  {title} {Mapping the presence of \textit{Wolbachia pipientis} on the
  phylogeny of filarial nematodes: evidence for symbiont loss during
  evolution},\ }\href@noop {} {\bibfield  {journal} {\bibinfo  {journal} {Int.
  J. Parasitol.}\ }\textbf {\bibinfo {volume} {34}},\ \bibinfo {pages} {191}
  (\bibinfo {year} {2004})}\BibitemShut {NoStop}%
\bibitem [{\citenamefont {LaSarre}\ \emph {et~al.}(2017)\citenamefont
  {LaSarre}, \citenamefont {McCully}, \citenamefont {Lennon},\ and\
  \citenamefont {McKinlay}}]{lasarre:2017}%
  \BibitemOpen
  \bibfield  {author} {\bibinfo {author} {\bibfnamefont {B.}~\bibnamefont
  {LaSarre}}, \bibinfo {author} {\bibfnamefont {A.~L.}\ \bibnamefont
  {McCully}}, \bibinfo {author} {\bibfnamefont {J.~T.}\ \bibnamefont
  {Lennon}},\ and\ \bibinfo {author} {\bibfnamefont {J.~B.}\ \bibnamefont
  {McKinlay}},\ }\bibfield  {title} {\bibinfo {title} {Microbial mutualism
  dynamics governed by dose-dependent toxicity of cross-fed nutrients},\
  }\href@noop {} {\bibfield  {journal} {\bibinfo  {journal} {ISME Journal}\
  }\textbf {\bibinfo {volume} {11}},\ \bibinfo {pages} {337} (\bibinfo {year}
  {2017})}\BibitemShut {NoStop}%
\bibitem [{\citenamefont {Schultz}\ and\ \citenamefont
  {Brady}(2008)}]{schultz:2008}%
  \BibitemOpen
  \bibfield  {author} {\bibinfo {author} {\bibfnamefont {T.~R.}\ \bibnamefont
  {Schultz}}\ and\ \bibinfo {author} {\bibfnamefont {S.~G.}\ \bibnamefont
  {Brady}},\ }\bibfield  {title} {\bibinfo {title} {Major evolutionary
  transitions in ant agriculture},\ }\href@noop {} {\bibfield  {journal}
  {\bibinfo  {journal} {PNAS}\ }\textbf {\bibinfo {volume} {105}},\ \bibinfo
  {pages} {5435} (\bibinfo {year} {2008})}\BibitemShut {NoStop}%
\bibitem [{\citenamefont {Currie}\ \emph {et~al.}(2003)\citenamefont {Currie},
  \citenamefont {Wong}, \citenamefont {Stuart}, \citenamefont {Schultz},
  \citenamefont {Rehner}, \citenamefont {Mueller}, \citenamefont {Sung},
  \citenamefont {Spatafora},\ and\ \citenamefont {Straus}}]{currie:2003}%
  \BibitemOpen
  \bibfield  {author} {\bibinfo {author} {\bibfnamefont {C.~R.}\ \bibnamefont
  {Currie}}, \bibinfo {author} {\bibfnamefont {B.}~\bibnamefont {Wong}},
  \bibinfo {author} {\bibfnamefont {A.~E.}\ \bibnamefont {Stuart}}, \bibinfo
  {author} {\bibfnamefont {T.~R.}\ \bibnamefont {Schultz}}, \bibinfo {author}
  {\bibfnamefont {S.~A.}\ \bibnamefont {Rehner}}, \bibinfo {author}
  {\bibfnamefont {U.~G.}\ \bibnamefont {Mueller}}, \bibinfo {author}
  {\bibfnamefont {G.-H.}\ \bibnamefont {Sung}}, \bibinfo {author}
  {\bibfnamefont {J.~W.}\ \bibnamefont {Spatafora}},\ and\ \bibinfo {author}
  {\bibfnamefont {N.~A.}\ \bibnamefont {Straus}},\ }\bibfield  {title}
  {\bibinfo {title} {Ancient tripartite coevolution in the attine ant-microbe
  symbiosis},\ }\href@noop {} {\bibfield  {journal} {\bibinfo  {journal}
  {Science}\ }\textbf {\bibinfo {volume} {299}},\ \bibinfo {pages} {386}
  (\bibinfo {year} {2003})}\BibitemShut {NoStop}%
\bibitem [{\citenamefont {Mueller}\ \emph {et~al.}(2001)\citenamefont
  {Mueller}, \citenamefont {Schultz}, \citenamefont {Currie}, \citenamefont
  {Adams},\ and\ \citenamefont {Malloch}}]{mueller:2001}%
  \BibitemOpen
  \bibfield  {author} {\bibinfo {author} {\bibfnamefont {U.~G.}\ \bibnamefont
  {Mueller}}, \bibinfo {author} {\bibfnamefont {T.~R.}\ \bibnamefont
  {Schultz}}, \bibinfo {author} {\bibfnamefont {C.~R.}\ \bibnamefont {Currie}},
  \bibinfo {author} {\bibfnamefont {R.~M.~M.}\ \bibnamefont {Adams}},\ and\
  \bibinfo {author} {\bibfnamefont {D.}~\bibnamefont {Malloch}},\ }\bibfield
  {title} {\bibinfo {title} {The origin of the attine ant-fungus mutualism},\
  }\href@noop {} {\bibfield  {journal} {\bibinfo  {journal} {Q Rev Biol}\
  }\textbf {\bibinfo {volume} {76}},\ \bibinfo {pages} {169} (\bibinfo {year}
  {2001})}\BibitemShut {NoStop}%
\bibitem [{\citenamefont {Hern\'andez}(1998)}]{hernandez:1998}%
  \BibitemOpen
  \bibfield  {author} {\bibinfo {author} {\bibfnamefont {M.~J.}\ \bibnamefont
  {Hern\'andez}},\ }\bibfield  {title} {\bibinfo {title} {Dynamics of
  transitions between population interactions: {A} nonlinear interaction
  $\alpha$-function defined},\ }\href@noop {} {\bibfield  {journal} {\bibinfo
  {journal} {P. Roy. Soc. B-Biol. Sci.}\ }\textbf {\bibinfo {volume} {265}},\
  \bibinfo {pages} {1433} (\bibinfo {year} {1998})}\BibitemShut {NoStop}%
\end{thebibliography}%


%

\end{document}